\documentclass[journal]{elsarticle}
\usepackage[paperwidth=18.4cm,paperheight=26cm,top=1.5cm,bottom=2cm,right=2cm]{geometry}
\usepackage{amssymb}
\usepackage{CJKutf8}
\usepackage{amsmath}
\usepackage{hyperref} 
\hypersetup{
colorlinks=true,
linkcolor=blue,
filecolor=blue,
urlcolor==blue,
citecolor=blue,
}
\biboptions{sort&compress} 
\usepackage{color}
\usepackage{setspace}
\usepackage[noend]{algpseudocode}
\usepackage{graphicx}
\usepackage{float} 
\usepackage{subfig}
\usepackage{algorithmicx,algorithm}
\usepackage{multirow}
\usepackage{bbm}
\def\degree{${}^{\circ}$}

\journal{Engineering Applications of Computational Fluid Mechanics}

\begin{document}
\begin{CJK}{UTF8}{gbsn}

\begin{frontmatter}

\title{QLingNet: An efficient and flexible modeling framework for subsonic airfoils}

\cortext[mycorrespondingauthor]{Corresponding author}

\author[nwpu_address]{Kuijun Zuo}

\author[nwpu_address]{Zhengyin Ye}

\author[cardc_address]{Linyang Zhu}

\author[cardc_address]{Xianxu Yuan \corref{mycorrespondingauthor}}
\ead{yuanxianxu2023@163.com}

\author[nwpu_address]{Weiwei Zhang \corref{mycorrespondingauthor}}
\ead{aeroelastic@nwpu.edu.cn}

\address[nwpu_address]{School of Aeronautics, Northwestern Polytechnical University, Xi'an, 710072, China}

\address[cardc_address]{State Key Laboratory of Aerodynamics, China Aerodynamics Research and Development Center, Mian'yang, 621000, China}

\begin{abstract}
{ \indent 
Artificial intelligence techniques are considered an effective means to accelerate flow field simulations. However, current deep learning methods struggle to achieve generalization to flow field resolutions while ensuring computational efficiency.
This paper presents a deep learning approach for rapid prediction of two types of subsonic flow fields with different resolutions.
Unlike convolutional neural networks, the constructed feature extractor integrates features of different spatial scales along the channel dimension, reducing the sensitivity of the deep learning model to resolution while improving computational efficiency.
Additionally, to ensure consistency between the input and output resolutions of the deep learning model, a memory pooling strategy is proposed, which ensures accurate reconstruction of flow fields at any resolution.
By conducting extensive qualitative and quantitative analyses on a given test dataset, it is demonstrated that the proposed deep learning model can achieve a three-order-of-magnitude speedup compared to CPU-based solvers while adapting to flow fields of arbitrary resolutions.
Moreover, the prediction accuracy for pressure exceeds 99\%, laying the foundation for the development of large-scale models in the field of aerodynamics.

}
\end{abstract}


\end{frontmatter}

\section{Introduction}
In the engineering applications of computational fluid dynamics (CFD) \cite{cutrone2024transition, chaiyanupong2024design}, the Reynolds-Averaged Navier-Stokes (RANS) method is widely employed for solving flow fields \cite{SIMSEK2023114298} and analyzing the aerodynamic performance of airfoils \cite{esfahanian2024aerodynamic, salimipour2024moving, XIE2023121002}.
However, as the complexity of engineering problems increases, solving the Navier-Stokes (NS) equations becomes time-consuming and requires significant memory resources \cite{zhong2024fast}.
Especially in the aerodynamic optimization design of aircraft \cite{shukla2024deep, yetkin2024investigation}, modifying a certain parameter often necessitates repetitive tasks such as grid partitioning and solving the NS equations. 
The flourishing development of artificial intelligence technology \cite{wang2024amsc, soler2024reinforcement, ismael2021deep, zhu2024vistfc} has provided a new perspective for the rapid solution of flow fields.

As a pioneering effort, Guo et al. \cite{guo2016convolutional} utilized convolutional neural networks (CNNs) to predict steady-state laminar velocity fields in both two and three dimensions. 
Through experiments, they found that CNNs could achieve computational speeds approximately four orders of magnitude faster than CPU-based solvers and two orders of magnitude faster than GPU-based solvers.
Wu and his team \cite{wu2022fast} took the signed distance field (SDF) and flow field boundary conditions as inputs to a neural network. 
They employed CNN to predict the velocity and pressure fields of the NACA0012 airfoil series under incompressible steady-flow conditions. 
Experimental results indicated that using neural networks could achieve a threefold increase in speed compared to traditional CFD simulation methods, with a prediction error of less than $1\%$.
As mentioned above, CNNs have been widely applied in the rapid simulation of airfoil flow fields. 
Similar works include the DeepCFD network architecture proposed by Mateus et al. \cite{ribeiro2020deepcfd}, the Mesh-Conv proposed by Hu et al. \cite{hu2022mesh}, and the CNNFOIL model for rapid simulation of transonic airfoil flow fields introduced by Cihat \cite{duru2021cnnfoil, duru2022deep}, among others.
In addition to CNNs, fully connected neural networks (FNNs) \cite{sun2021deep, shukla2024deep, leer2021fast}, generative adversarial networks (GANs) \cite{haizhou2022generative, wu2020deep, wang2023general}, graph neural networks (GNNs) \cite{yang2022amgnet, ogoke2021graph, li2022integrated}, point clouds \cite{chen2024pointgpt, abbas2022geometrical}, and Transformer neural network architectures \cite{jiang2023transcfd, hemmasian2023reduced, deng2023prediction, zuo2023fast} have also been extensively employed in the rapid simulation and solution tasks of flow fields.
However, a survey of existing literature reveals challenges in achieving predictions for flow fields at different resolutions. 
The main challenges include: 
(1) Current neural network models lack effective methods for handling multi-scale flow field structures.
(2) If we consider the flow field prediction task as a field-to-field modeling problem similar to a black-box model, ensuring adaptive handling of flow field resolution during both model training and inference stages is essential. 
It is also crucial to maintain consistency between the predicted output resolution of the neural network model and the input resolution.
(3) While multi-layer perceptrons (MLP) can generalize across different grid resolutions, the training cost and memory overhead of the model exhibit a quadratic relationship with the input resolution of the airfoil grid.
Inspired by the work of Chen et al. \cite{chen2107cyclemlp}, we propose the QLingNet (see \ref{name} for an explanation of the name), a deep learning flow field rapid prediction network architecture designed for multi-scale subsonic airfoil flow field structures.
The primary contributions of this study can be summarized as follows:
\begin{itemize}
	\item Proposed a computationally efficient and flexible deep learning model for the rapid prediction of subsonic variable-topology airfoil flow fields.
	
	\item Generated two types of subsonic airfoil flow fields with distinct topological structures, namely the University of Illinois Urbana-Champaign (UIUC) \cite{swannet2024towards} airfoil flow field database and the class function/shape function transformation (CST) \cite{lane2010inverse} airfoil parameterized perturbation airfoil flow field database.
	
	\item Utilizing the QLingNet ensures linear computational complexity for calculating flow fields at different resolutions.
	
	\item Integrated a memory pool module into the QLingNet, ensuring that the neural network's output flow field size remains consistent with the input flow field size.
\end{itemize}

The rest of this paper is organized as follows. 
Section \uppercase\expandafter{\romannumeral2} 
mainly describes the physical governing equations used for flow field simulation and the QLingNet network framework.
Section \uppercase\expandafter{\romannumeral3} primarily discusses the two types of airfoil flow field datasets with different geometries and resolutions.
Section \uppercase\expandafter{\romannumeral4} shows and discusses the results of the QLingNet neural network model training and test. 
And the conclusion is given in Section \uppercase\expandafter{\romannumeral5}.

\section{Methodology}
\subsection{Physical equations}

The training data for the neural network model used here is obtained through simulation calculations conducted by the Platform for Hybrid Engineering Simulation of Flows (PHengLEI) \cite{zhao2020design} software developed by the China Aerodynamics Research and Development Center (CARDC).
Specifically, the PHengLEI computation program obtains the physical information at each discrete point in the flow field by solving the RANS equations on the structural grid of a two-dimensional airfoil.

The airfoil is simulated under the following conditions: Mach number (Ma) = 0.5, angle of attack (AOA) = 3.86 \degree, and Reynolds number (Re) = $3 \times 10^6$.
Below is a brief introduction to the governing equations and the SA (Spalart-Allmaras) turbulence model employed in the simulation software calculations.
The RANS equations under Favre averaging are:

\begin{equation} \label{eq:1}
\frac{\partial \bar{\rho}}{\partial t}+\frac{\partial}{\partial x_j}\left(\bar{\rho} \bar{u_j}\right)=0
\end{equation}

\begin{equation} \label{eq:2}
\frac{\partial}{\partial t}\left(\bar{\rho} \tilde{u}_i\right)+\frac{\partial}{\partial x_j}\left(\bar{\rho} \tilde{u}_i \tilde{u}_j\right)=-\frac{\partial \bar{p}}{\partial x_i}+\frac{\partial \bar{\tau}_{i j}}{\partial x_j}-\frac{\partial}{\partial x_j} \overline{\rho u_i^{\prime \prime} u_j^{\prime \prime}}
\end{equation}

\begin{equation}
	\frac{\partial}{\partial t}(\bar{\rho} \tilde{e})+\frac{\partial}{\partial x_j}\left(\bar{\rho}\tilde{u_j} \tilde{e}\right)=\frac{\partial}{\partial x_j}\left(\kappa \frac{\partial \bar{T}}{\partial x_j}\right)-\bar{p} \frac{\partial \tilde{u}_j}{\partial x_j}+\bar{\Phi}-\overline{u_j^{\prime \prime} \frac{\partial p^{\prime}}{\partial x_j}}-\frac{\partial}{\partial x_j} \overline{\rho e^{\prime \prime} u_j^{\prime \prime}}
\end{equation}

In above equations, $t$ represents time, $p$ and $\rho$ denote pressure and density, $u_i$ and $u_j$ represent the velocity components in the $x_i$ and $x_j$ directions, respectively. 
$\tau_{ij}$ represents the viscous stress tensor.
The symbol "$\sim$" denotes the Favre-averaged physical quantity, a prime superscript indicates the time-averaged fluctuating momentum, and a double prime superscript signifies the fluctuating momentum under Favre averaging.
$\overline{\tau_{ij}}$ represents the mean viscous stress, $\overline{\rho u_i^{\prime \prime} u_j^{\prime \prime}}$ represents the Reynolds stress, $\bar{\Phi}$ denotes the mean flow viscous dissipation rate, the work done by fluctuating pressure along fluctuating displacement is represented by $\overline{u_j^{\prime\prime}\frac{\partial p^{\prime}}{\partial x_j}}$, and the correlation between fluctuating energy and fluctuating velocity is denoted by $\overline{\rho e^{\prime\prime}u_j^{\prime\prime}}$.

Here, the SA turbulence model is employed to close the RANS equations. Turbulent eddy viscosity is defined as:

\begin{equation} 
	\mu_{\mathrm{t}}=\rho\hat{v}f_{v1}
\end{equation}

\begin{equation} \label{eq:5}
	f_{v1}=\frac{\chi^{3}}{\chi^{3}+C_{v1}^{3}},\chi=\frac{\hat{\nu}}{\nu}
\end{equation}

In Eq. \ref{eq:5}, $C_{\nu1}$ is a model parameter typically set to 7.1, $\nu$ is the kinematic molecular viscosity.
Further deriving the transport equation: 

\begin{equation} \label{eq:6}
	 \begin{aligned}
	 \frac{\partial\hat{v}}{\partial t}+u_{j}\frac{\partial\hat{v}}{\partial x_{j}}& =C_{b1}(1-f_{t2})\hat{S}\hat{v}-\left(C_{w1}f_{w}-\frac{C_{b1}}{\kappa^{2}}f_{t2}\right)\left(\frac{\hat{v}}{d^{\prime}}\right)^{2}  \\
	 &+\frac{1}{\sigma}\left\{\frac{\partial}{\partial x_{j}}\left[(v+\hat{v})\frac{\partial\hat{v}}{\partial x_{j}}\right]+C_{b2}\frac{\partial\hat{v}}{\partial x_{i}}\cdot\frac{\partial\hat{v}}{\partial x_{i}}\right\}
	 \end{aligned}
\end{equation}

\begin{equation} \label{eq:7}
	 \hat{S}=S+\frac{\hat{v}}{\kappa^{2}d^{2}}f_{v2}
\end{equation}

\begin{equation} \label{eq:8}
	f_{v2}=1-\frac{\chi}{1+\chi f_{v1}},f_{t2}=C_{t3}\mathrm{exp}(-C_{t4}\chi^{2}),f_{w}=g\left(\frac{1+C_{w3}^{6}}{g^{6}+C_{w3}^{6}}\right)^{1/6}
\end{equation}

\begin{equation} \label{eq:9}
	g=r+C_{w2}(r^6-r),r=\frac{\hat{v}}{\hat{S}\kappa^2d^2},C_{w1}=\frac{C_{b1}}{\kappa^2}+\frac{1+C_{b2}}{\sigma}
\end{equation}

In Eq. \ref{eq:6}, \ref{eq:7}, \ref{eq:8}, \ref{eq:9}, $\bar{v}$ is the state variable of the SA turbulence model, $d$ is the wall distance, $S$ represents the magnitude of the vorticity. 
$C_{b1}$, $C_{b2}$, $\sigma$, $C_{w1}$, $C_{w2}$, $C_{w3}$, $C_{v1}$, $\kappa$, $C_{t3}$ and $C_{t4}$ represent model parameters.
In accordance with the empirical findings outlined in reference \cite{zhao2020design}, the model parameters are conventionally set to $C_{b1}=0.1355$, $C_{b2}=0.622$, $\sigma=\frac{2}{3}$, $\kappa=0.41$, $C_{w2}=0.3$, $C_{w3}=2$.

\subsection{Cycle fully-connected module} \label{Cycle FC}

\begin{figure*}[!h]
	\begin{center}
		\includegraphics[width=1 \linewidth]{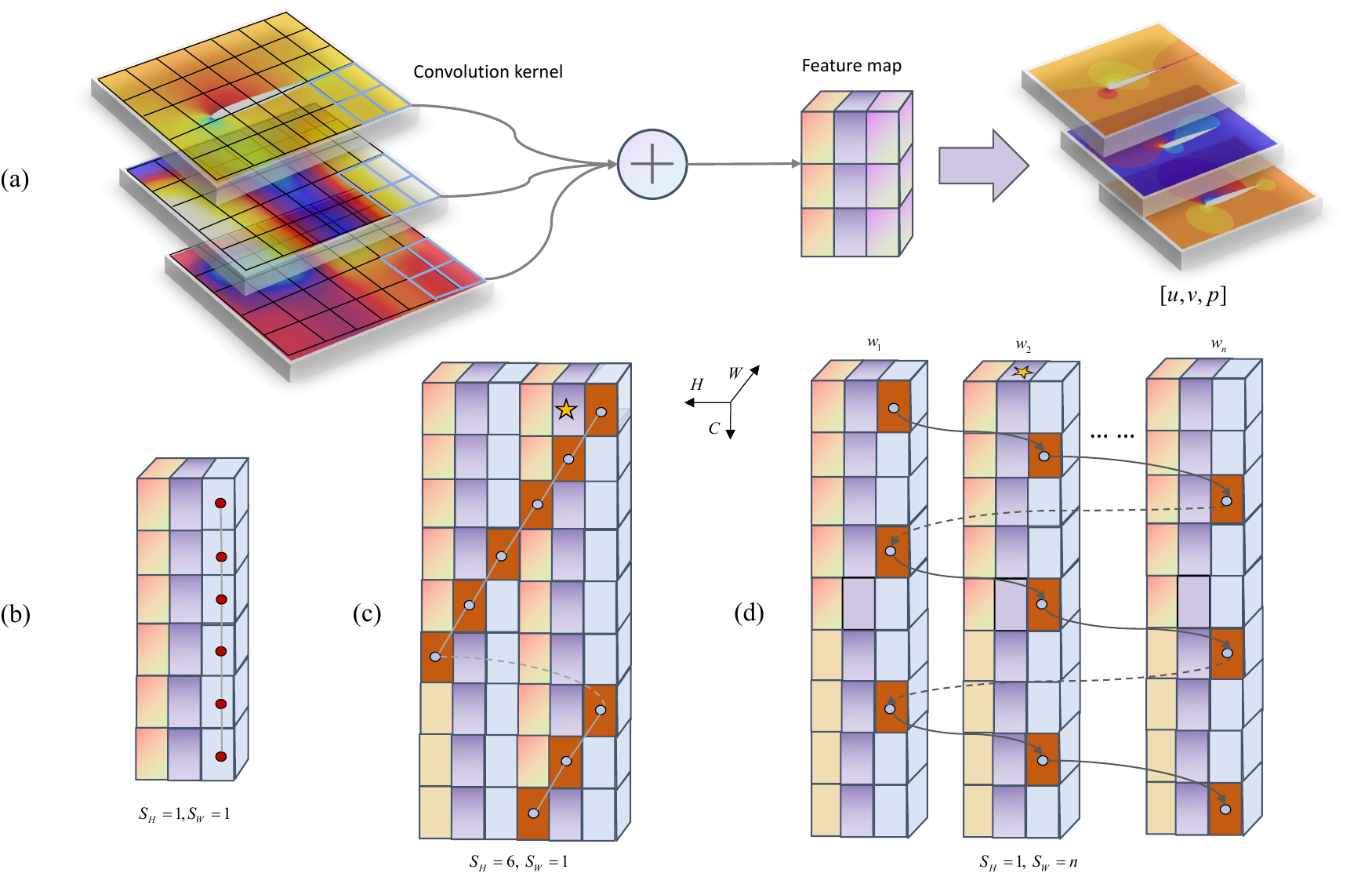}
	\end{center}  \vspace{-2mm}  
	\caption{{{Comparison of CNN and CycleFC feature extraction methods.}
	}} \label{cycleMLP} 
\end{figure*}

First, we represent the flow field feature maps input to the neural network as $X\in\mathbb{R}^{H{\times}W{\times}C_{in}}$, where $H$, $W$, and $C_{in}$ represent the height, width, and number of channels of the grid, respectively.
As shown in Figure \ref{cycleMLP}(a), for convolutional neural networks, the main approach is to extract flow field features by sliding fixed convolutional kernels over the given feature maps. The calculation formula is as follows:

\begin{equation}
	\left.\left\{\begin{aligned}&H_{out}=\frac{H_{in}+2\times P_0-K_0}{S_0}+1,\\&W_{out}=\frac{W_{in}+2\times P_1-K_1}{S_1}+1.\end{aligned}\right.\right.
\end{equation}

In the above formula, $P_{i}$ represents the padding size, $K_i$ represents the size of the convolutional kernel, and $S_i$ represents the stride of the convolutional kernel.
However, the above-mentioned spatial convolutional neural network architectures struggle to flexibly handle flow fields with different resolutions.
The channel fully connected (channel FC) model shown in Figure \ref{cycleMLP}(b) extracts flow field features along the channel dimension at fixed positions $(i, j)$.
Although it can flexibly handle flow fields with different resolutions, but it lacks the capability to capture contextual information of the flow field.

As shown in Figure \ref{cycleMLP}(c) and Figure \ref{cycleMLP}(d), Cycle fully-connected (Cycle FC) introduces the $step size (S_H, S_W)$ on the basis of the channel FC model to increase the model's receptive field. 
Its calculation formula is as follows:
\begin{equation} \label{eq:11}
	\mathrm{CycleFC}(\boldsymbol{X})_{i,j,:}=\sum_{c=0}^{C_{i\boldsymbol{n}}}\boldsymbol{X}_{i+\delta_i(c),j+\delta_j(c),c}\cdot\boldsymbol{W}_{c,:}^{\mathbf{mlp}}+\boldsymbol{b}
\end{equation}

In Eq. (\ref{eq:11}), $\boldsymbol{W^\mathrm{mlp}}\in\mathbb{R}^{C_{in}\times C_{out}}$ and $\boldsymbol{b} \in\mathbb{R}^{C_{out}}$ are the model parameters to be optimized. 
$\delta_{i}(c)$ and $\delta_{j}(c)$ represent the offsets along the channels $c$ in the $S_H$ and $S_W$ directions, respectively, and they are defined as:

\begin{equation}
	\delta_i(c)=(c\bmod S_H)-1, \delta_j(c)=(\lfloor\frac{c}{S_H}\rfloor\bmod{S_W})-1
\end{equation}

For Figure \ref{cycleMLP}(c), when $S_H=6$, $S_W=1$, with the pentagram in the figure as the reference coordinate, when $c=\{0,1,2,\ldots,5\}$, the offset in the $S_H$ direction is $\delta_i(c)=\{-1,0,1,2,3,4\}$.
In Figure \ref{cycleMLP}(d), when $S_H=1$ and $S_W=n$, $\delta_j(c)=\{-1,0,1,...,(n-2)\}$.
The derivation above reveals that the Cycle FC module incorporates offset terms along the channel dimension, allowing it to retain the efficient computational capabilities while also integrating contextual information from different spatial points. 
Moreover, this network architecture can generalize well to flow fields of different resolutions.

\subsection{QLingNet architecture} 

\begin{figure*}[!h]
	\begin{center}
		\includegraphics[width=1 \linewidth]{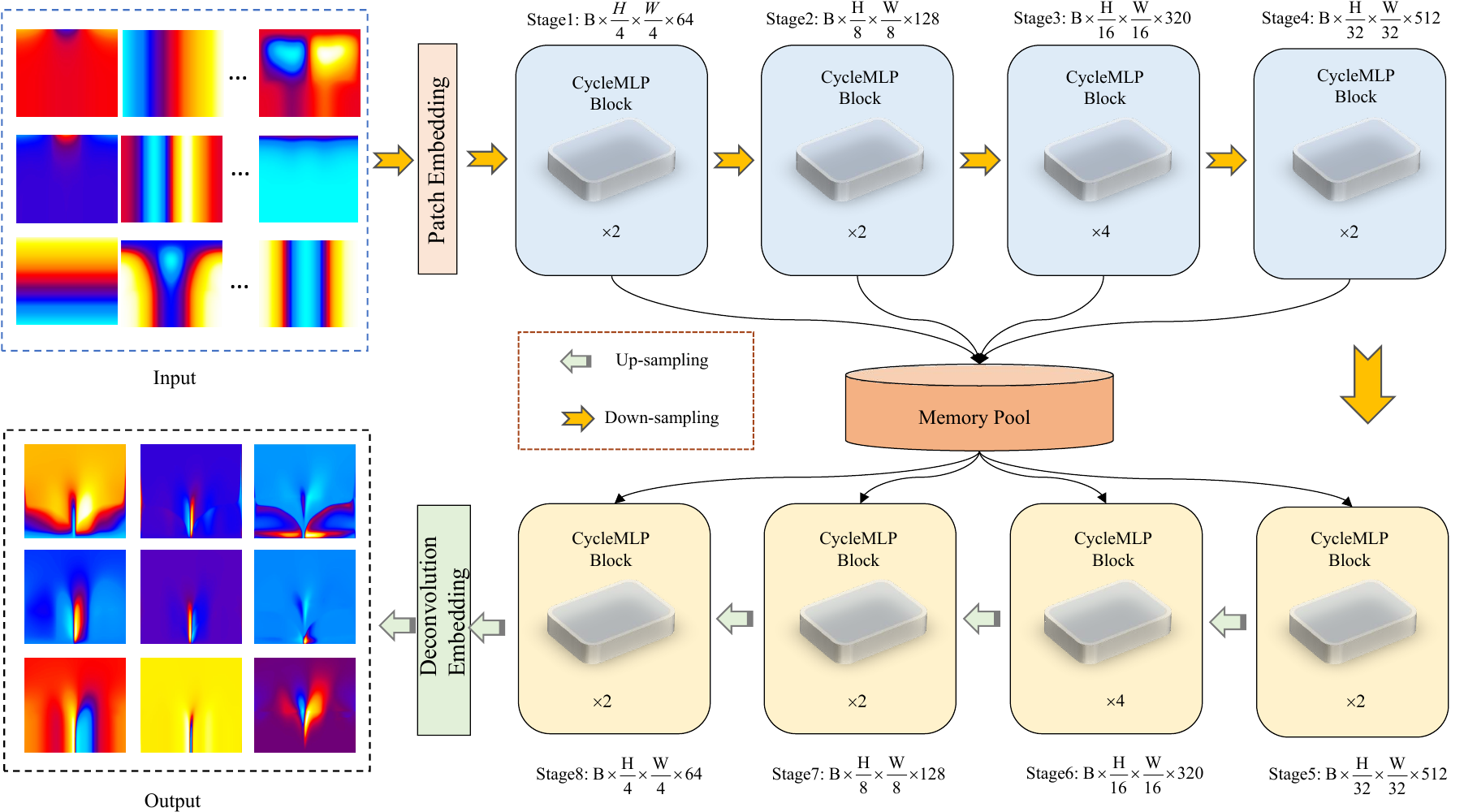}
	\end{center}  \vspace{-2mm}  
	\caption{{{QLingNet neural network architecture.}
	}} \label{QLingNet} 
\end{figure*}

Based on the Cycle FC component described in Section \ref{Cycle FC}, we construct the QLingNet. 
Below, we provide an introduction to the fundamental modules involved in Figure \ref{QLingNet}.

\textbf{Input:} The flow field data for the QLingNet comprises two types. 
The first type consists of flow field data obtained through PHengLEI computations based on the UIUC airfoil, with a resolution of $55 \times 403$. 
The second type is derived from the NACA0012 airfoil using the CST parameterization method, resulting in a flow field resolution of $120 \times 364$.
Both types of flow fields have 13 channels ($x, y, x_0, y_0, \xi, \eta, SDF, M_x, M_y, \frac{\partial x}{\partial \xi}, \frac{\partial x}{\partial \eta}, \frac{\partial y}{\partial \xi}, \frac{\partial y}{\partial \eta}$). 
The interpretation of the above parameters can be found in Section \ref{data}.
Therefore, the flow field resolutions inputted into QLingNet are $55 \times 403 \times 13$ and $120 \times 364 \times 13$, respectively.

\textbf{Patch Embedding:} This module segments the flow field data inputted into QLingNet into a series of patches. 
Specifically, it mainly uses a two-dimensional convolutional module to accomplish the segmentation task, with hyperparameters convolutional kernel=7, stride=4, padding=2.
Furthermore, it maps the channels to a higher dimension through linear mapping. 
Therefore, the dimension of the input features after passing through patch embedding module is $\frac{H}{4} \times \frac{W}{4} \times 64$.

\begin{figure*}[!h]
	\begin{center}
		\includegraphics[width=1 \linewidth]{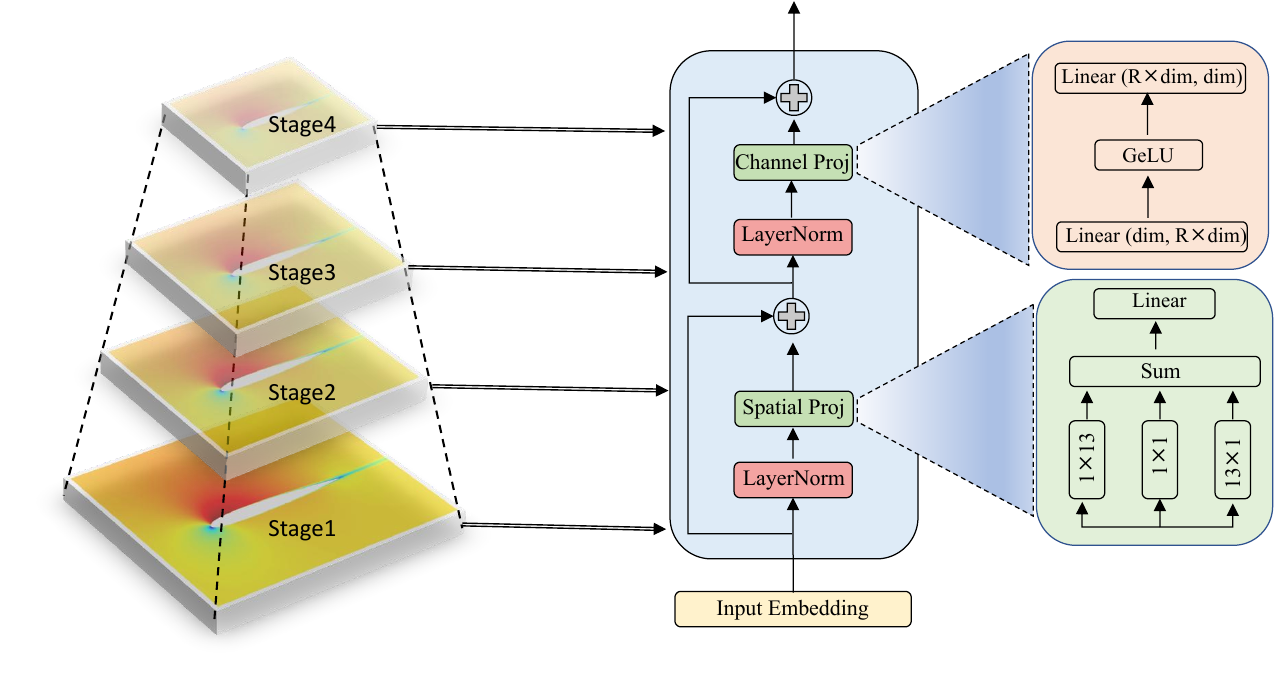}
	\end{center}  \vspace{-2mm}  
	\caption{{{Feature pyramid and CycleMLP block.}
	}} \label{CycleMLP Block} 
\end{figure*}

\textbf{CycleMLP Block:} 
The main body of QLingNet consists of a series of CycleMLP blocks, where multiple CycleMLP blocks extract spatial features of the flow field at different scales in a pyramid structure and perform information fusion. The detailed network structure of the CycleMLP module is illustrated in Figure \ref{CycleMLP Block}.
The spatial mapping is performed by three parallel CycleFC modules, each with different $stepsize$ ($1 \times13, 1 \times 1, 13 \times 1$). 
The information extracted by these three CycleFC modules is further fused using an attention layer. 
Channel mapping is achieved through two linear layers followed by the GeLU activation function. 
LayerNorm and residual network layers are added before and after both the spatial and channel mappings, respectively.

\textbf{Stage:} Each stage consists of multiple CycleMLP blocks stacked together. 
In QLingNet, both the encoding and decoding layers are composed of 4 stages. 
The feature map resolution processed within the same stage is consistent. 
After each stage, there is a down-sampling layer, which reduces the spatial dimensions of the feature maps while increasing the number of channels. 
This ensures a reduction in computational complexity while operating in a high-dimensional feature space.

\textbf{Memory pool:} To ensure that QLingNet can adapt to flow fields of different resolutions, a memory pool is designed to store the feature map resolutions at each stage of the encoding down-sampling phase. 
During the decoding up-sampling phase, the model sequentially retrieves the corresponding feature resolutions from the memory pool, ensuring that the final output resolution of the flow field matches the original resolution. 
Here, the output channel number is 9 ($u,v,p,\frac{\partial u}{\partial\xi},\frac{\partial u}{\partial\eta},\frac{\partial v}{\partial\xi},\frac{\partial v}{\partial\eta},\frac{\partial p}{\partial\xi},\frac{\partial p}{\partial\eta}$), representing the velocity field, pressure field, and their corresponding gradient information in the computational coordinates.

\textbf{Deconvolution Embedding:} To restore the feature map resolution of the embedded patches back to the original flow field size, a deconvolution embedding layer is added to the final layer of the model. 
After passing through this layer, the output size of the flow field feature map becomes $H \times W \times 9$.

\section{Data preparation} \label{data}

\begin{figure*}[!h]
	\begin{center}
		\includegraphics[width=0.9 \linewidth]{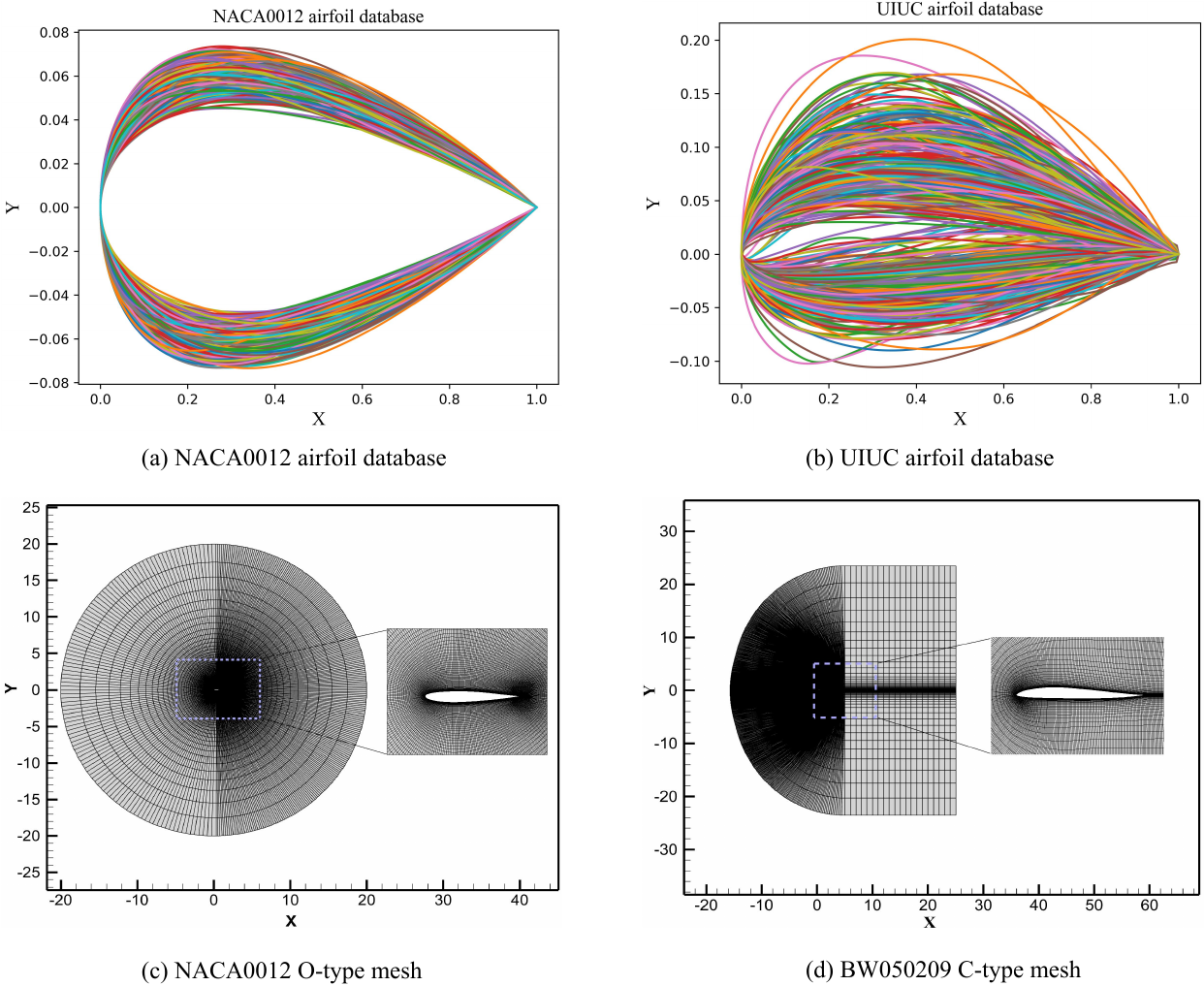}
	\end{center}  \vspace{-2mm}  
	\caption{{{Two types of grids with different topological structures and airfoil database.}
	}} \label{mesh} 
\end{figure*}

Figure \ref{mesh} presents two types of flow field data with different topological structures, and they are utilized to test the generalization capability of QLingNet when facing flow fields with different resolutions.
As shown in Figure \ref{mesh}(a), the first type of test case involves 500 airfoil geometries generated by perturbing the NACA0012 airfoil using the CST parameterization method. 
The grid is partitioned using an O-type mesh, as shown in Figure \ref{mesh}(c), with a grid size of $120 \times 364$, where the grid height is 120 and the number of grid points per layer is 364.
As depicted in Figure \ref{mesh}(b), the second type of test case consists of 270 airfoil geometries with significant variations from the UIUC public airfoil database. 
The grid is primarily partitioned using a C-type mesh, as shown in Figure \ref{mesh}(d), with a grid size of $55 \times 403$.
To facilitate subsequent data analysis, the first dataset will be named 'NACA0012-CST' and the second dataset will be named 'UIUC'.
Using the PHengLEI solver to obtain training data for the neural network, the operating conditions are Re=$3 \times 10^6$, AOA=3.86\degree, and Ma=0.5.
$80\%$ of the aforementioned flow field data is utilized for training the QLingNet model, $10\%$ is allocated for cross-validation, and the remaining $10\%$ is designated for testing.

\begin{figure*}[!h]
	\begin{center}
		\includegraphics[width=1 \linewidth]{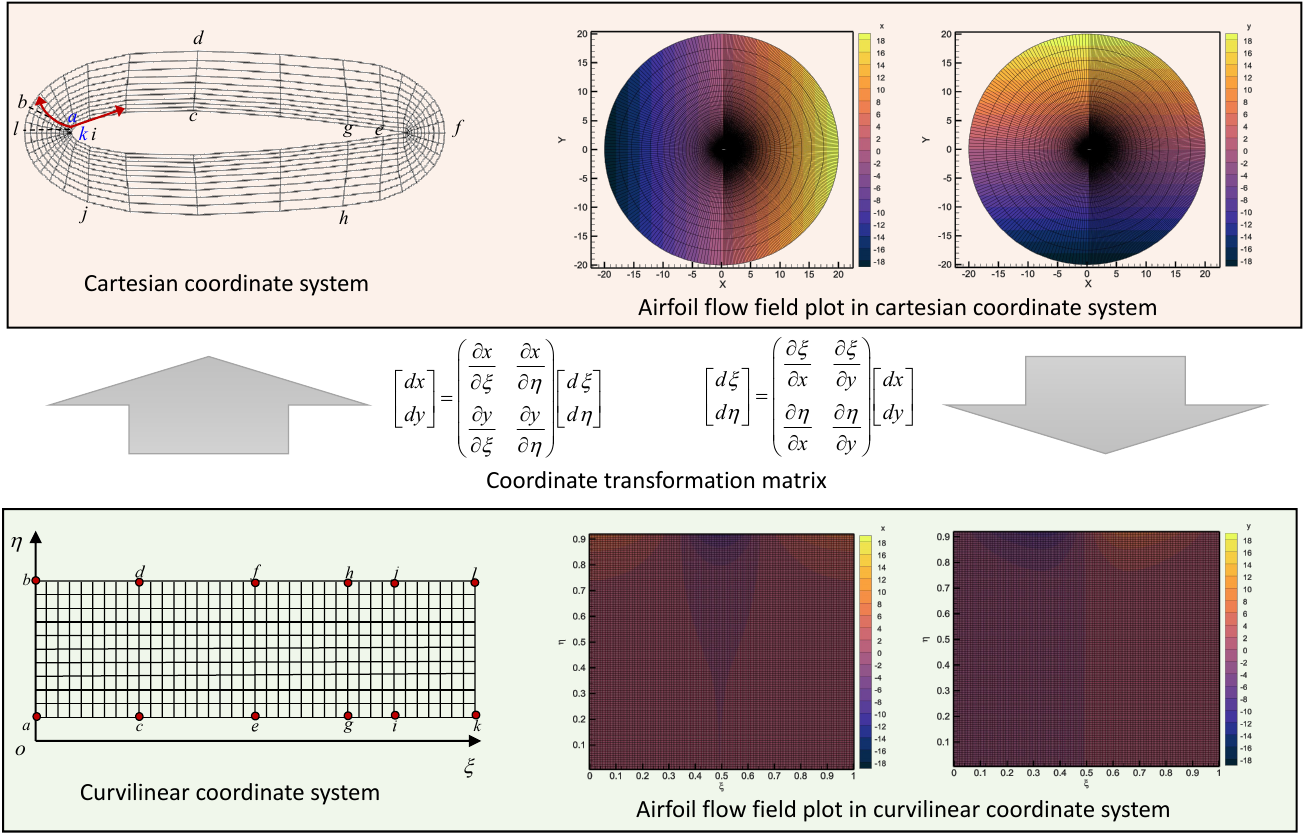}
	\end{center}  \vspace{-2mm}  
	\caption{{{Conversion between curvilinear coordinates and Cartesian coordinates.}
	}} \label{coordinate transformation} 
\end{figure*}
%
The mapping function for the QLingNet neural network, as a type of end-to-end black-box model, can be represented as follows:
\begin{equation} \label{eq:13}
	f(x, y, x_0, y_0, \xi, \eta, SDF, M_x, M_y, \frac{\partial x}{\partial \xi}, \frac{\partial x}{\partial \eta}, \frac{\partial y}{\partial \xi}, \frac{\partial y}{\partial \eta},) = (u, v, p, \frac{\partial u}{\partial \xi}, \frac{\partial u}{\partial \eta}, \frac{\partial v}{\partial \xi}, \frac{\partial v}{\partial \eta}, \frac{\partial p}{\partial \xi}, \frac{\partial p}{\partial \eta})
\end{equation}

On the left side of Eq. (\ref{eq:13}) are the thirteen features input to the neural network, while on the right side are the nine flow field variables to be predicted.
As shown in Figure \ref{coordinate transformation}, in order to facilitate the subsequent training of the QLingNet model, following the processing strategy in reference \cite{zuo2024fast}, the flow field data is transformed from Cartesian coordinates ($x, y$) to curvilinear coordinates ($\xi, \eta$).
$x_0$ and $y_0$ are the coordinates of the first layer of grid points on the surface. 
$\xi$ and $\eta$ represent the curvilinear coordinates, calculated by the following formula:
$\xi=(i-1)/(i_{max}-1),\eta=(j-1)/(j_{max}-1)$, where $i$ and $j$ represent the indices in different directions of the grid, and $i_{max}, j_{max}$ are the maximum values of the grid nodes.
Due to the rich flow field information near the boundary, it is the area of ​​greater concern throughout the simulation process. 
Therefore, we have designed three features: $SDF$, $M_x$, and $M_y$. 
Their calculation formulas are as follows:
\begin{equation} \label{eq:14}
	\begin{cases}M=e^{-|SDF|},\Psi_M(x)=M\times x,\Psi_M(y)=M\times y,\\SDF(i,j)=\min_{(i^*,j^*)\in Z}|(i,j)-(i^*,j^*)|sign[f(i,j)]\end{cases}
\end{equation}

$SDF$ is used to describe the signed distance from a flow field point to the surface of the airfoil.
Otherwise, the purpose of features $M_x$ and $M_y$ is to artificially reweight each grid point in the flow field based on the $SDF$ parameter. 
Grid points closer to the airfoil surface are given greater weight, while those farther away are given smaller weight, as the numerical values of the flow field in the far field are precisely the areas we are less concerned about.
As shown in Figure \ref{coordinate transformation}, $\frac{\partial x}{\partial \xi}$, $\frac{\partial x}{\partial \eta}$, $\frac{\partial y}{\partial \xi}$, and $\frac{\partial y}{\partial \eta}$ represent the Jacobian matrices used in the coordinate transformation.
The right-hand side of Eq. (\ref{eq:14}) represents the predictions of QLingNet, which are the velocity field and pressure field. 
Since flow parameters often vary significantly at the leading and trailing edges of the airfoil, the gradient information of velocity ($\frac{\partial u}{\partial \xi}, \frac{\partial u}{\partial \eta}, \frac{\partial v}{\partial \xi}, \frac{\partial v}{\partial \eta}$) and pressure ($\frac{\partial p}{\partial \xi}, \frac{\partial p}{\partial \eta}$) is added as constraint terms to the loss function. 
The loss function used during model training is the mean squared error (MSE), calculated as:

\begin{equation} \label{eq: 15}
	\begin{gathered}
	MSE_{loss}=\frac{1}{9\times N}\sum_{i=1}^{N}[(u_{i}^{t}-u_{i}^{k})^{2}+(v_{i}^{t}-v_{i}^{k})^{2}+(p_{i}^{t}-p_{i}^{k})^{2}+ \\
	((\frac{\partial u}{\partial\xi})_i^t-(\frac{\partial u}{\partial\xi})_i^k)^2+((\frac{\partial u}{\partial\eta})_i^t-(\frac{\partial u}{\partial\eta})_i^k)^2+((\frac{\partial v}{\partial\xi})_i^t-(\frac{\partial v}{\partial\xi})_i^k)^2+ \\
	((\frac{\partial v}{\partial\eta})_{i}^{t}-(\frac{\partial v}{\partial\eta})_{i}^{k})^{2}+((\frac{\partial p}{\partial\xi})_{i}^{t}-(\frac{\partial p}{\partial\xi})_{i}^{k})^{2}+((\frac{\partial p}{\partial\eta})_{i}^{t}-(\frac{\partial p}{\partial\eta})_{i}^{k})^{2}] 
	\end{gathered}
\end{equation}

Here, $N$ is the number of samples, $\wp_i^{t}$ and $\wp_i^{k}$ $(\wp:u,v,p,\frac{\partial u}{\partial\xi},\frac{\partial u}{\partial\eta},\frac{\partial v}{\partial\xi},\frac{\partial v}{\partial\eta},\frac{\partial p}{\partial\xi},\frac{\partial p}{\partial\eta})$ represent the ground truth and predicted values by QLingNet, respectively.

\section{Results and discussions}

\subsection{Ablation experiments}

Firstly, ablative experiments were conducted on the model to select the optimal hyperparameters. 
The initial learning rate during the training process was set to $5 \times 10^{-5}$. 
The Adam optimizer was employed to optimize the model parameters during training, with a batch size of 1. 
The program code was written using the Python language and the PyTorch deep learning framework. 
Additionally, the training of the QLingNet model was accelerated by utilizing an RTX 3090 GPU. 
The MSE mentioned in Section \ref{data} was utilized as the loss function during model training.
Additionally, to facilitate model convergence, a learning rate scheduler was employed to automatically adjust the learning rate during the training process. 
Here, the learning rate step size was set to 50, and the hyperparameter $gamma$ was set to 0.1, indicating that every 50 iterations during model training, the learning rate was multiplied by 0.1.
The model completed a total of 500 iterations of training.

\begin{figure*}[!h]
	\begin{center}
		\includegraphics[width=1 \linewidth]{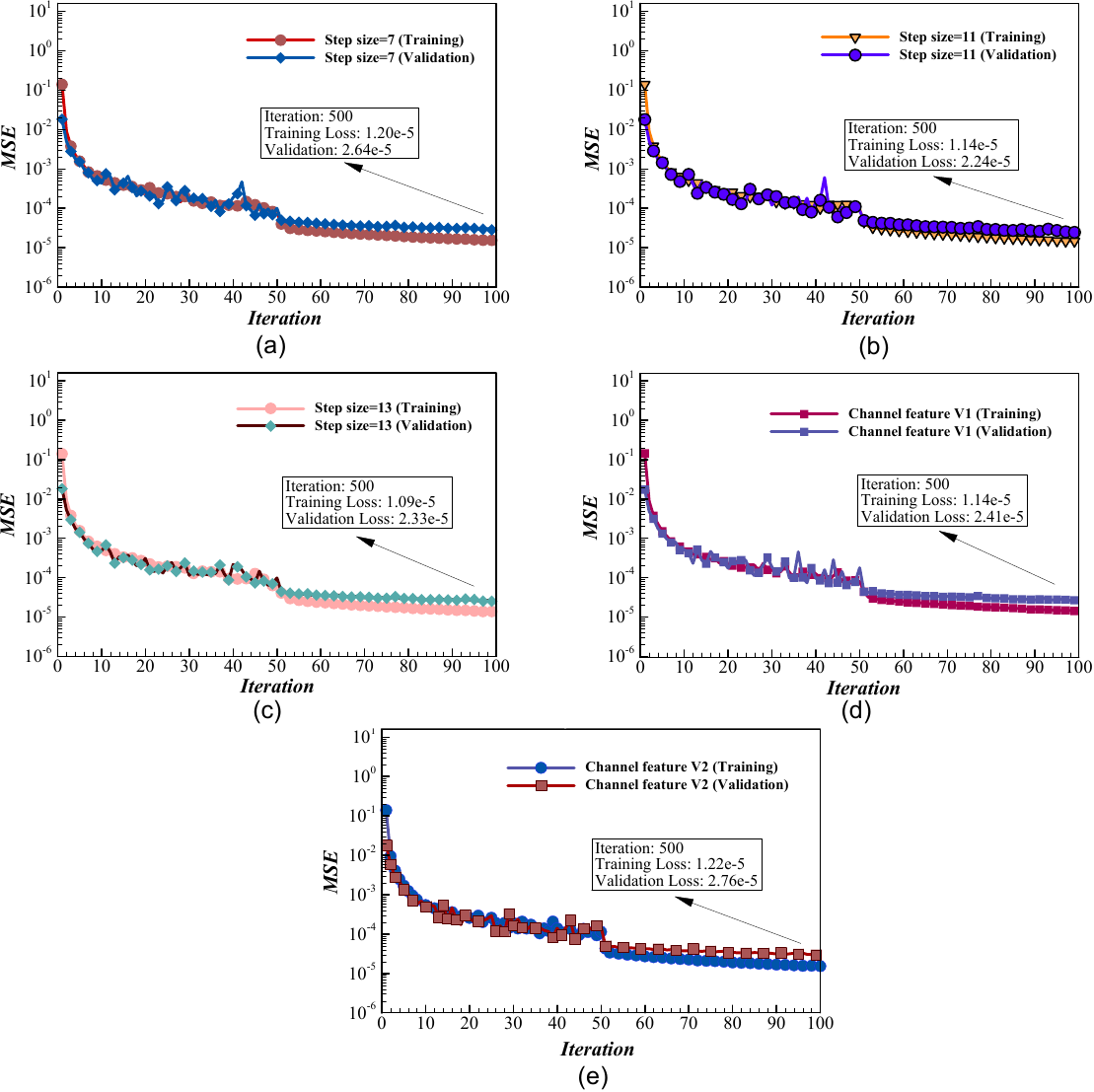}
	\end{center}  \vspace{-2mm}  
	\caption{{{Loss function variation curves of QLingNet trained with different hyperparameters.}
	}} \label{loss} 
\end{figure*}

Figure \ref{loss} and Table \ref{loss table} respectively present the variation curves of the loss function during training under five different hyperparameter settings, along with detailed parameter configuration information. 
As indicated in the second column of Table \ref{loss table}, the first hyperparameter is the $stepsize$ of CycleFC mentioned in Section \ref{Cycle FC}. 
A larger value of this parameter implies a larger receptive field of the model. 
However, from the training results, a larger receptive field does not necessarily correspond to better generalization performance. 
From Model 1 to Model 2, with the increase in the $stepsize$ value, the loss function slightly decreases. 
However, when this value increases from 11 to 13, although a smaller loss function is achieved on the training set, the loss function on the cross-validation set increases. Therefore, in this study, the $stepsize$ is set to 11.

\begin{table}[tb]
	\caption{Ablation experiment results of QLingNet} 
	\label{loss table}
	\begin{center}
		\begin{tabular}{cccccc}
			\hline \hline
			Name &
			Stepsize &
			Layers &
			Embedding dimension &
			Training loss &
			Validation loss \\ \hline 
			Model 1 &
			7 &
			\begin{tabular}[c]{@{}c@{}}{[}2, 2, 4, 2{]}\\ {[}2, 4, 2, 2{]}\end{tabular} &
			\begin{tabular}[c]{@{}c@{}}{[}64, 128, 320, 512{]}\\ {[}512, 320, 128, 64{]}\end{tabular} &
			$1.20 \times 10^{-5}$ &
			$2.64 \times 10^{-5}$ \\
			Model 2 &
			11 &
			\begin{tabular}[c]{@{}c@{}}{[}2, 2, 4, 2{]}\\ {[}2, 4, 2, 2{]}\end{tabular} &
			\begin{tabular}[c]{@{}c@{}}{[}64, 128, 320, 512{]}\\ {[}512, 320, 128, 64{]}\end{tabular} &
			$1.14 \times 10^{-5}$ &
			$2.24 \times 10^{-5}$ \\
			Model 3 &
			13 &
			\begin{tabular}[c]{@{}c@{}}{[}2, 2, 4, 2{]}\\ {[}2, 4, 2, 2{]}\end{tabular} &
			\begin{tabular}[c]{@{}c@{}}{[}64, 128, 320, 512{]}\\ {[}512, 320, 128, 64{]}\end{tabular} &
			$1.09 \times 10^{-5}$ &
			$2.33 \times 10^{-5}$ \\
			Model 4 &
			13 &
			\begin{tabular}[c]{@{}c@{}}{[}2, 2, 4, 2{]}\\ {[}2, 4, 2, 2{]}\end{tabular} &
			\begin{tabular}[c]{@{}c@{}}{[}64, 128, 256, 512{]}\\ {[}512, 256, 128, 64{]}\end{tabular} &
			$1.14 \times 10^{-5}$ &
			$2.41 \times 10^{-5}$ \\
			Model 5 &
			13 &
			\begin{tabular}[c]{@{}c@{}}{[}4, 4, 6, 4{]}\\ {[}4, 6, 4, 4{]}\end{tabular} &
			\begin{tabular}[c]{@{}c@{}}{[}64, 128, 256, 512{]}\\ {[}512, 256, 128, 64{]}\end{tabular} &
			$1.22 \times 10^{-5}$ &
			$2.76 \times 10^{-5}$ \\ \hline
		\end{tabular}
	\end{center} 
	\vspace{-1.5em}
\end{table} 

As shown in the third column of Table \ref{loss table}, the second hyperparameter is "layers", which represents the number of CycleMLP Block modules in each stage of the QLingNet network. 
Similarly, from the test results of Model 4 to Model 5, it can be observed that as the number of CycleMLP Blocks increases in each stage, the loss function on both the training set and the cross-validation set increases. 
Therefore, in this study, the layers in the encoding and decoding stages of the model are set to 2, 2, 4, 2, and 2, 4, 2, 2, respectively.
As shown in the fourth column of Table \ref{loss table}, the third hyperparameter is "embedding dimension", which corresponds to the "layers" hyperparameter one by one. 
It represents the size of the feature dimension in each stage during the down-sampling or up-sampling process of the model. 
By comparing the test results of Model 3 and Model 4, it can be observed that this parameter also affects the test results. 
According to the test results, in the decoding and encoding stages of the model, this value is set to 64, 128, 320, 512, and 512, 320, 128, 64, respectively.
Additionally, Figure \ref{loss} presents the curves of the loss function for the five different models on both the training set and the cross-validation set. 
At the initial training stage, the loss function curves of all five models show a decreasing trend with an increase in the number of iterations. 
However, as depicted in Figure \ref{loss}(b), corresponding to Model 2 in Table \ref{loss table}, the loss function on the cross-validation set is the smallest, indicating the best generalization performance of the model under the current parameters. 
Based on the above analysis, the parameters of Model 2 are selected as the hyperparameters for QLingNet during the model training process.

\subsection{Analysis of flow field prediction results}

\subsubsection{Analysis of NACA0012-CST dataset test results}

\begin{figure*}[!h]
	\begin{center}
		\includegraphics[width=1 \linewidth]{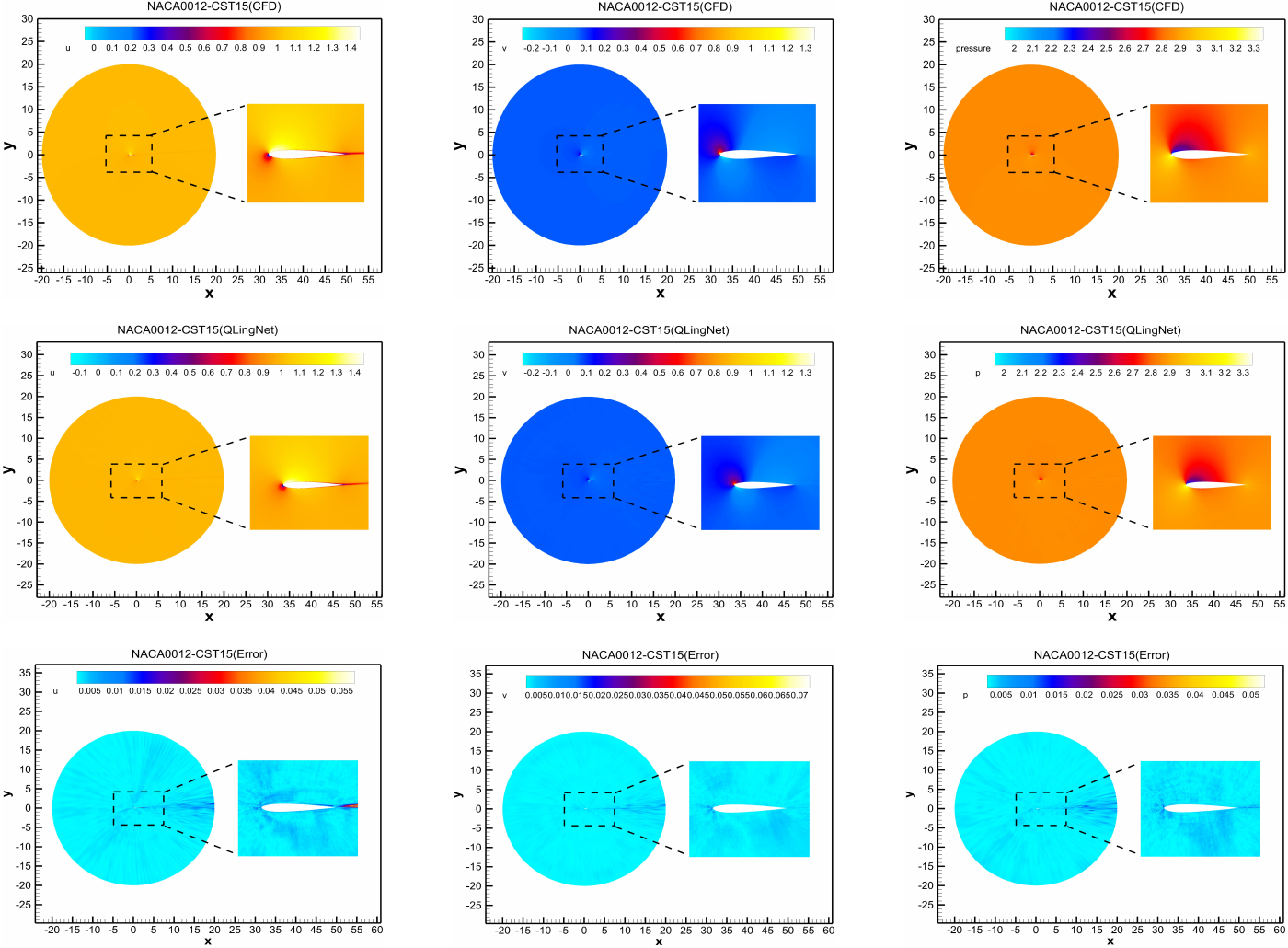}
	\end{center}  \vspace{-2mm}  
	\caption{{{Visualization comparing the CFD-calculated and neural network-predicted values of velocity ($u$, $v$) and pressure ($p$) for the NACA0012-CST15 airfoil, alongside corresponding absolute error plots.}
	}} \label{naca0012_cst_15} 
\end{figure*}

\begin{figure*}[!h]
	\begin{center}
		\includegraphics[width=1 \linewidth]{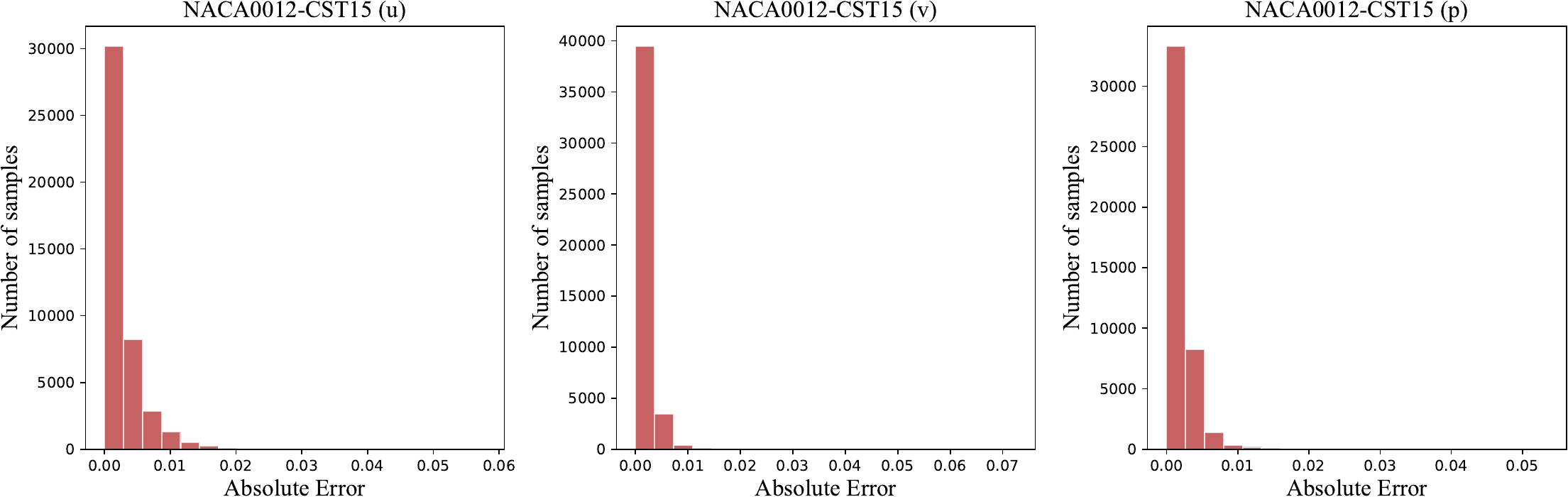}
	\end{center}  \vspace{-2mm}  
	\caption{{{Histogram of the absolute error distribution between CFD and QLingNet.}
	}} \label{naca0012_cst_15_error} 
\end{figure*}

We evaluate the flow field prediction performance of the QLingNet model using airfoils that were not included in the training dataset, as described in Section \ref{data}.
The initial testing focuses on the NACA0012-CST airfoil database, where 500 airfoils are encoded as NACA0012-CST$i$ ($i=1,2,...,500$). 
We specifically select the NACA0012-CST15 airfoil from the test set to evaluate the flow field prediction accuracy of the QLingNet model.
From Figure \ref{naca0012_cst_15}, it can be observed that there is good consistency between the CFD calculated results and QLingNet predictions.
Furthermore, the absolute error plots in Figure \ref{naca0012_cst_15} reveal that for velocity $u$, the error values range from $5 \times 10^{-3}$ to $5.5 \times 10^{-2}$; for velocity $v$, the absolute error values range from $5 \times 10^{-3}$ to $7 \times 10^{-2}$; and for pressure $p$, the absolute error values range from $5 \times 10^{-3}$ to $5 \times 10^{-2}$.
Otherwise, by examining the absolute error histograms in Figure \ref{naca0012_cst_15_error}, it is evident that the majority of errors for velocity $u$, $v$, and pressure $p$ are less than $1 \times 10^{-2}$. 
This indicates that the proposed QLingNet neural network model achieves a satisfactory prediction performance.

\begin{figure*}[!h]
	\begin{center}
		\includegraphics[width=1 \linewidth]{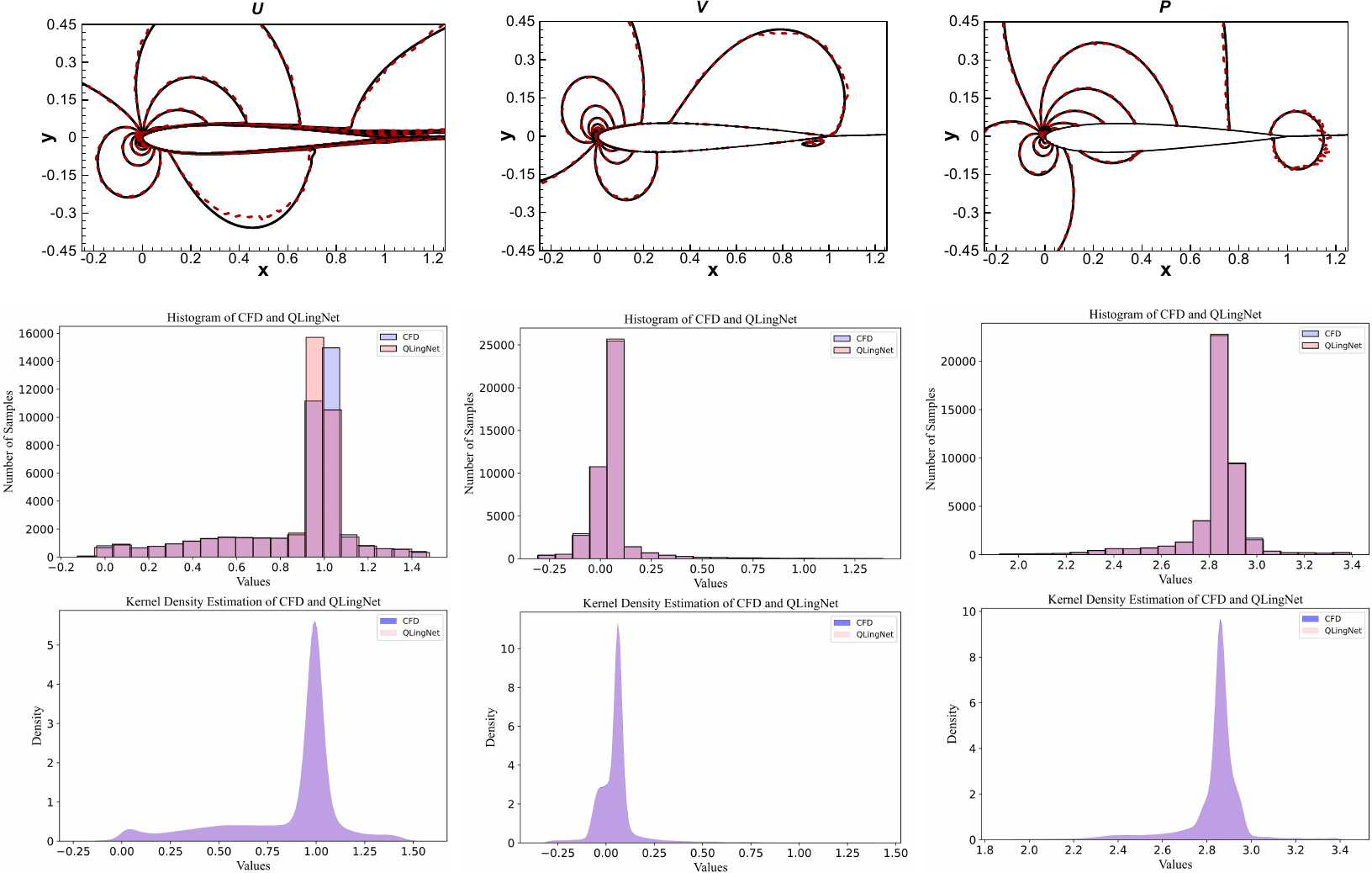}
	\end{center}  \vspace{-2mm}  
	\caption{{{First row: Contour plots between QLingNet flow field predictions and CFD computational results. The solid black line represents the CFD calculation values, while the dashed red line represents the QLingNet prediction results. Second row: Histogram comparing the data distribution between CFD and QLingNet. Third row: Kernel density plot comparing QLingNet and CFD.}
	}} \label{naca0012_cst_15_contour} 
\end{figure*}

To further test the accuracy of QLingNet's flow field prediction results, Figure \ref{naca0012_cst_15_contour} provides contour plots comparing CFD with QLingNet, histograms of data distribution, and corresponding kernel density plots.
From the test results in Figure \ref{naca0012_cst_15_contour}, the contour plots between CFD and QLingNet generally exhibit good fitting. However, there are certain areas where the fitting effect for velocity u is less ideal. 
This observation is also evident from the histograms of data distribution, where the majority of the data distribution shows good fitting, with noticeable differences only in the vicinity of the value 1. 

\begin{figure*}[!h]
	\begin{center}
		\includegraphics[width=1 \linewidth]{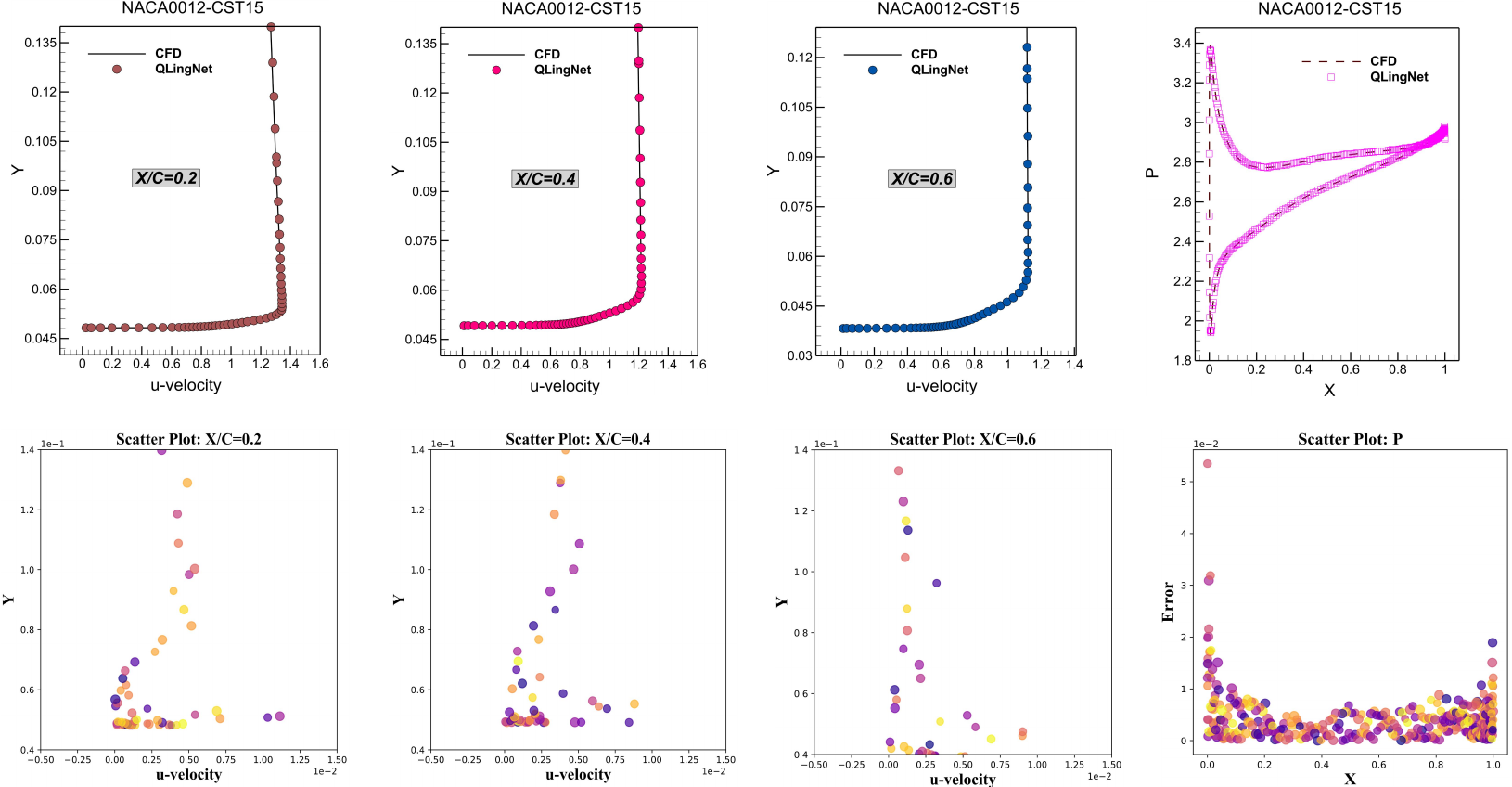}
	\end{center}  \vspace{-2mm}  
	\caption{{{First row: Velocity fitting curves and near-wall pressure fitting curves of CFD and QLingNet at different station points. Second row: Scatter plots of velocity absolute errors at different station points and near-wall pressure absolute error scatter plots between CFD and QLingNet.}
	}} \label{naca0012_cst_15_velocity} 
\end{figure*}

Nonetheless, further examination of the kernel density plots reveals minimal impact on the density plots due to subtle differences in the data. 
The higher peaks in the kernel density plot indicate denser data at those locations, which is consistent with the display results of the data distribution histogram.
Overall, the test results fully meet the engineering requirements.
For velocity v and pressure p, the fitting between CFD and QLingNet is relatively good in both contour plots and histograms of data distribution.
This indicates that as a black-box model, QLingNet is capable of training based on historical data to predict the flow field of new airfoils with different data distributions, achieving excellent predictive performance.

Figure \ref{naca0012_cst_15_velocity} depicts the velocity profile fitting curves between QLingNet and CFD at three different station points, as well as the pressure profile fitting curves at the airfoil surface, along with scatter plots of the absolute errors corresponding to each curve.
In Figure \ref{naca0012_cst_15_velocity}, the first three columns depict the velocity fitting curves of velocity u at positions 0.2, 0.4, and 0.6 on the airfoil boundary, respectively, comparing CFD with QLingNet. 
The test results demonstrate that at these three positions, both the CFD calculated results and the QLingNet predicted curves fit well.
Furthermore, from the corresponding scatter plots of absolute errors, it is observed that at position 0.2, most absolute error values are below $5 \times 10^{-3}$ with a maximum error of $1.25 \times 10^{-2}$, while at positions 0.4 and 0.6, the majority of absolute error values are below $7.5 \times 10^{-3}$, with only a few outliers ranging from $7.5 \times 10^{-3}$ to $1 \times 10^{-2}$.
Additionally, the last column of Figure \ref{naca0012_cst_15_velocity} illustrates the good fitting effect of pressure curves between QLingNet predictions and CFD calculations, with the majority of absolute error values being less than $1\times 10^{-2}$.
From the above test results, QLingNet achieved good predictive performance when facing the NACA0012-CST test data. 
Additionally, to comprehensively test the model's generalization capability, \ref{NACA0012 CST} provides the prediction results of more test cases.

\subsubsection{Analysis of test results for the UIUC dataset}

\begin{figure*}[!h]
	\begin{center}
		\includegraphics[width=1 \linewidth]{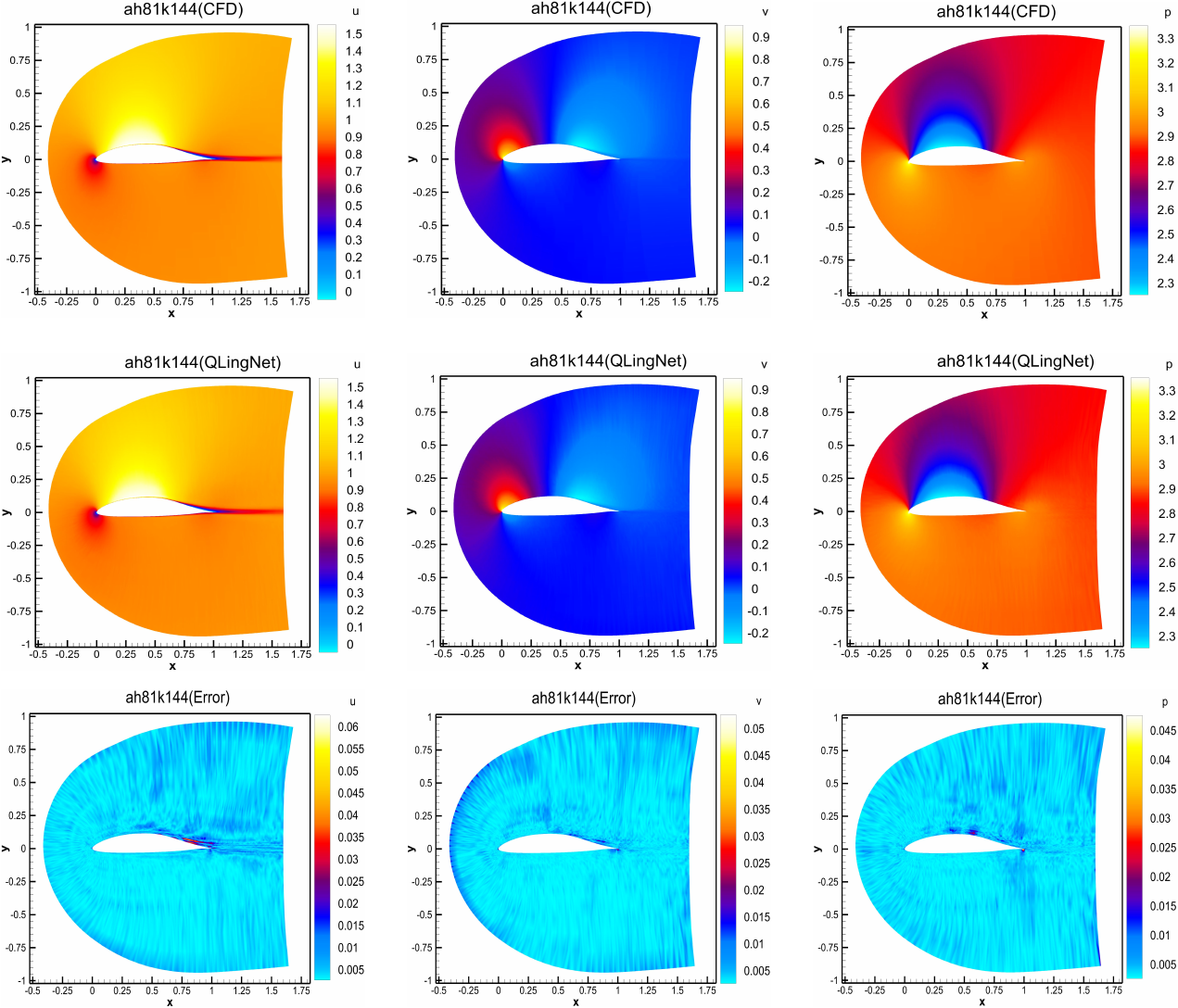}
	\end{center}  \vspace{-2mm}  
	\caption{{{Comparison between the CFD computational results and the predicted results of the QLingNet for ah81k144, alongside corresponding absolute error plots.}
	}} \label{ah81k144} 
\end{figure*}

\begin{figure*}[!h]
	\begin{center}
		\includegraphics[width=1 \linewidth]{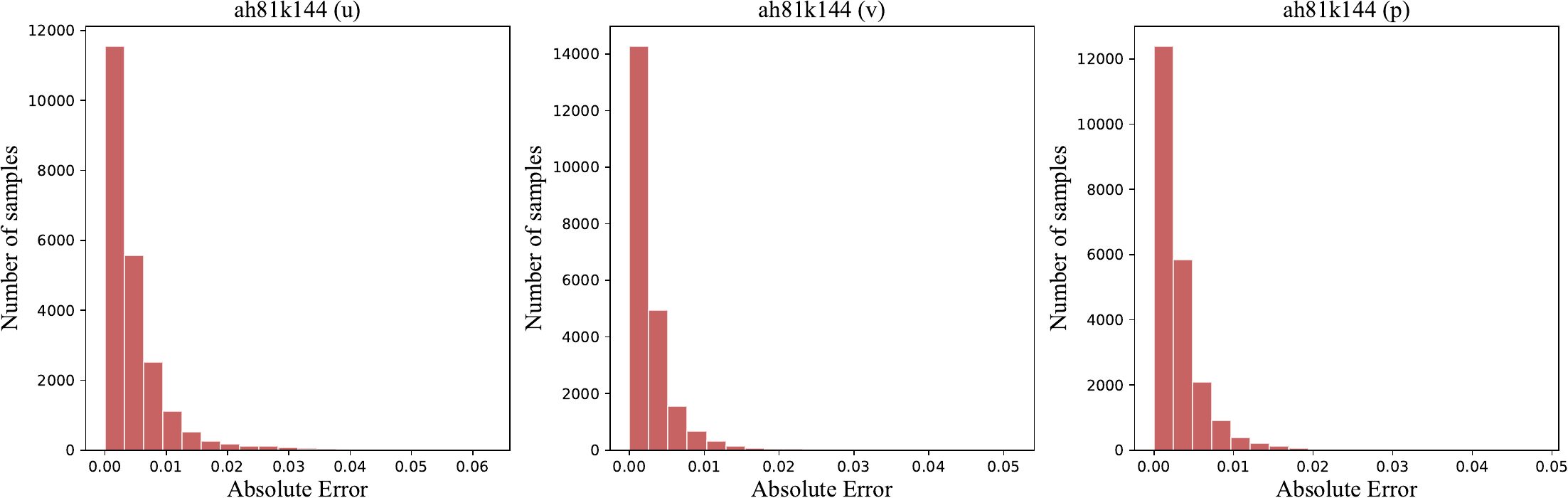}
	\end{center}  \vspace{-2mm}  
	\caption{{{Histogram of the absolute error distribution between CFD and QLingNet.}
	}} \label{ah81k144_error_hist} 
\end{figure*}

To assess the predictive accuracy and generalization capability of QLingNet across flow fields of different resolutions, Figure \ref{ah81k144} presents the flow field prediction results for the ah81k144 airfoil from the UIUC airfoil database.
Figure \ref{ah81k144} illustrates that the flow field predictions of QLingNet exhibit a high degree of similarity with the results of CFD calculations. 
Furthermore, from the contour plots of absolute errors between the predicted values of QLingNet and the CFD computed values, it can be observed that for the velocity component 
$u$, the absolute error ranges from $5 \times 10^{-3}$ to $6 \times 10^{-2}$, for the velocity component $v$, the absolute error ranges from $5 \times 10^{-3}$ to $5 \times 10^{-2}$, and for the pressure $p$, the absolute error ranges from $5 \times 10^{-3}$ to $4.5 \times 10^{-2}$.
Furthermore, from the histogram in Figure \ref{ah81k144_error_hist} depicting the distribution of absolute errors between CFD and QLingNet, it is evident that for both velocity components 
$u$ and $v$, as well as pressure $p$, the majority of error values are centered around zero.
Overall, the numerical range of errors lies between 0 and $2 \times 10^{-2}$, indicating a high level of predictive accuracy.

\begin{figure*}[!h]
	\begin{center}
	\includegraphics[width=1 \linewidth]{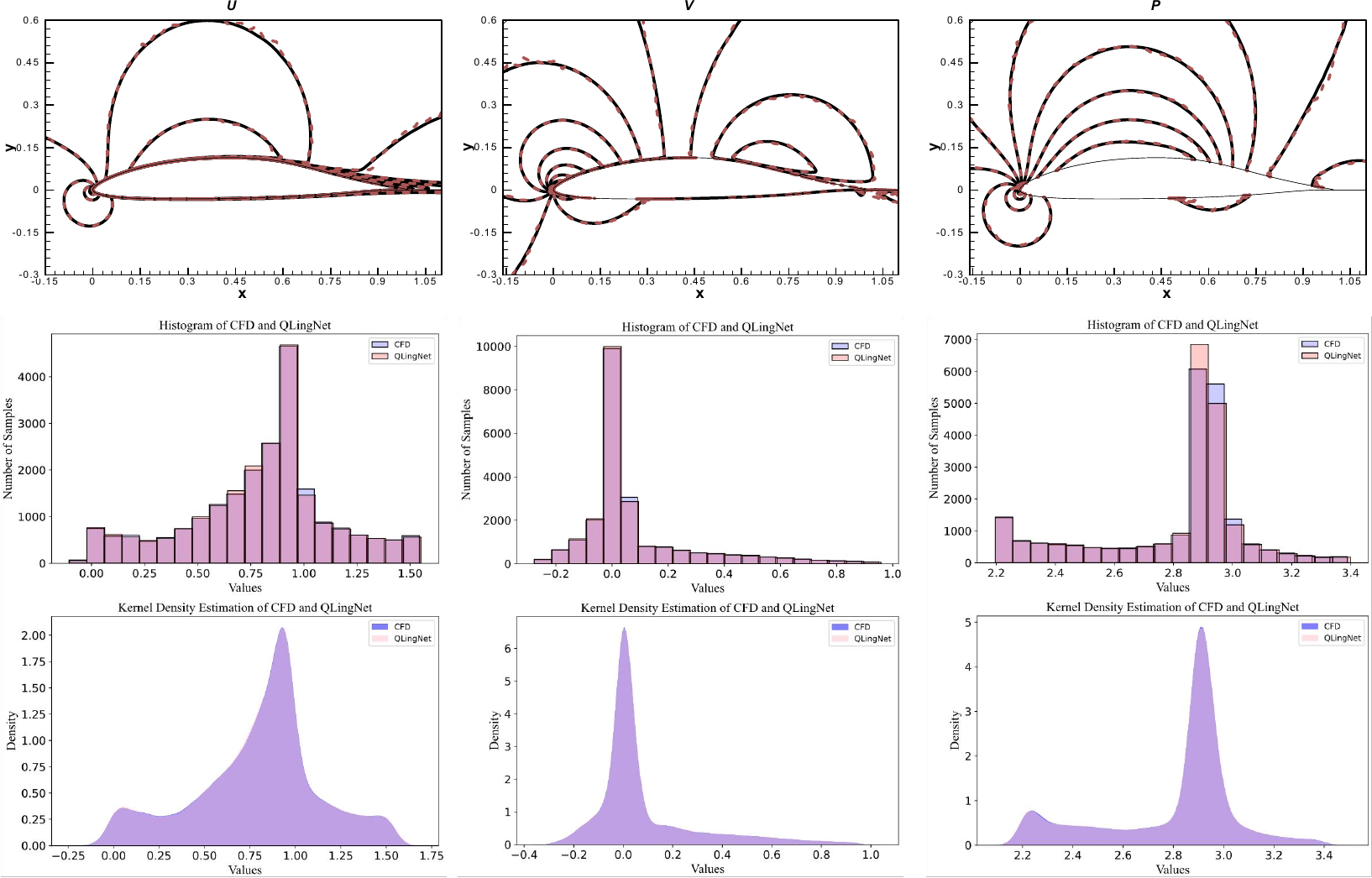}
	\end{center}  \vspace{-2mm}  
	\caption{{{First row: Contour plots between QLingNet flow field predictions and CFD computational results. The solid black line represents the CFD calculation values, while the dashed red line represents the QLingNet prediction results. Second row: Histogram comparing the data distribution between CFD and QLingNet. Third row: Kernel density plot comparing QLingNet and CFD..}
}} \label{ah81k144_hist} 
\end{figure*}

For a more detailed assessment of QLingNet's predictive accuracy, Figure \ref{ah81k144_hist} provides contour plots comparing predicted values to ground truth, histograms illustrating the distribution of flow field data, and corresponding kernel density plots.
From the first row of Figure \ref{ah81k144_hist}, it's evident that the lines between CFD and QLingNet fit perfectly. 
Additionally, the histograms in the second row show that, except for some differences in the histogram of pressure data, the distribution histograms of CFD computed data and QLingNet predicted data generally match well. 
Any differences in data distribution occur in adjacent regions, indicating close proximity of data in those locations. 
Furthermore, examining the kernel density plots in the third row reveals that the data distribution is dense around peaks of the curves, indicating similar trends in data distribution. 
Overall, even when facing flow fields of different resolutions, QLingNet demonstrates strong predictive capabilities.

\begin{figure*}[!h]
	\begin{center}
		\includegraphics[width=1 \linewidth]{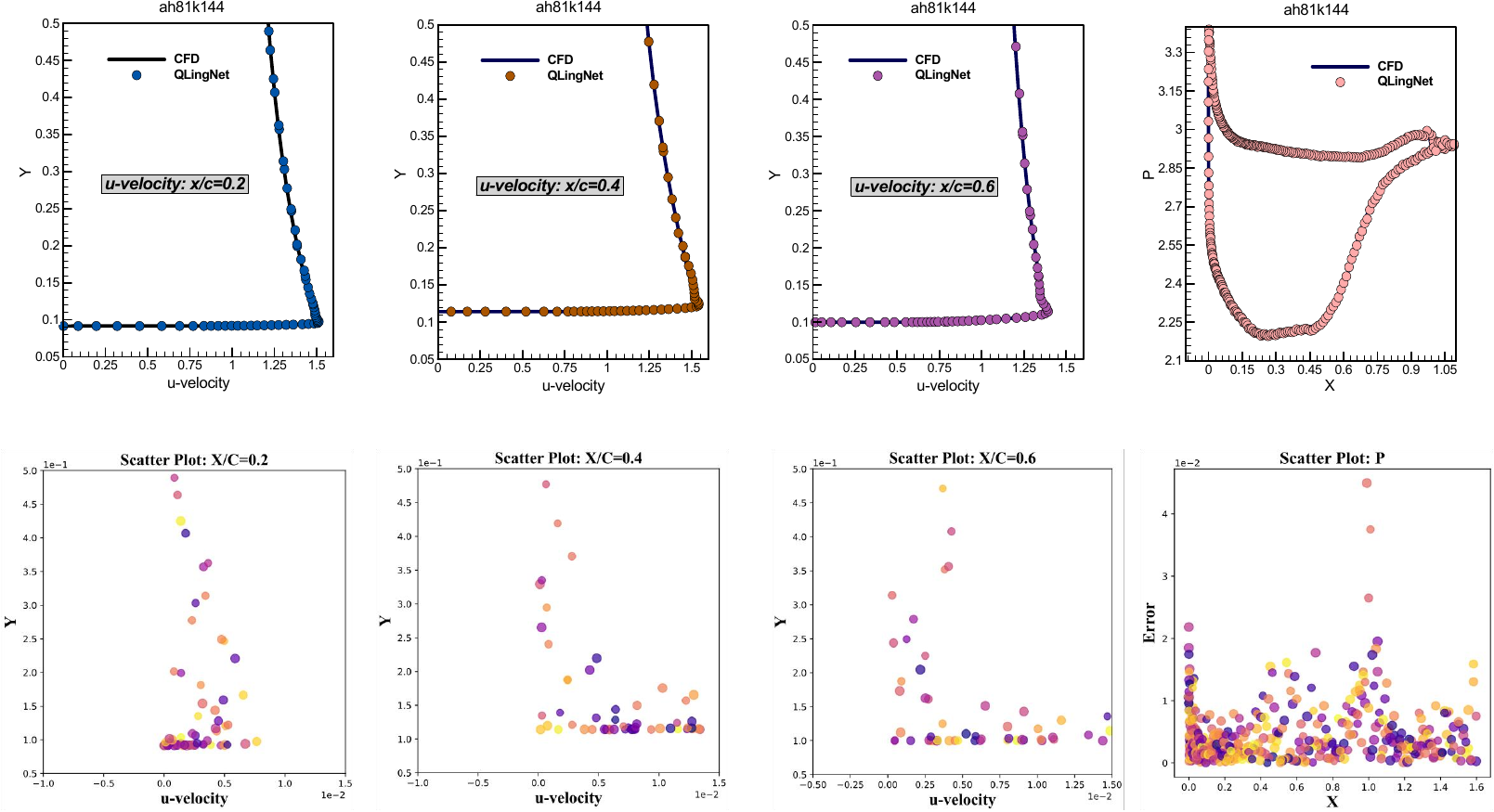}
	\end{center}  \vspace{-2mm}  
	\caption{{{First row: Velocity fitting curves and near-wall pressure fitting curves of CFD and QLingNet at different station points. Second row: Scatter plots of velocity absolute errors at different station points and near-wall pressure absolute error scatter plots between CFD and QLingNet.}
	}} \label{ah81k144_velocity_scatter} 
\end{figure*}

The above analyses have focused on QLingNet's predictive accuracy at the macroscopic scale of the flow field. 
Figure \ref{ah81k144_velocity_scatter} provides fitting curves of velocity $u$ at three different stations on the airfoil near-wall, as well as fitting curves of pressure distribution on the surface.
It can be observed that at stations 0.2, 0.4, and 0.6, the curves of CFD and QLingNet fit well, and similarly, for the pressure curve $p$, a good fitting effect is also achieved.
In the second row of Figure \ref{ah81k144_velocity_scatter}, scatter plots of absolute errors corresponding to each fitting curve are provided. 
For the velocity $u$ at station 0.2, the error data are mostly within the numerical range of $5 \times 10^{-3}$, with only some error points falling between $5 \times 10^{-3}$ and $1 \times 10^{-2}$. 
At station 0.4, the error data are primarily within the range of 0 to $1.5 \times 10^{-2}$.
Similarly, at station 0.6, the error data also fall within the range of 0 to $1.5 \times 10^{-2}$.
Regarding the error in pressure distribution on the surface, the majority of error data points are less than $2 \times 10^{-2}$, with only a few error data points falling between $2 \times 10^{-2}$ and $4 \times 10^{-2}$.
Table \ref{acceleratation ratio} provides comparisons between CFD computation time and QLingNet prediction time for more test airfoils.
It is evident that QLingNet prediction speed improves by three orders of magnitude compared to CFD simulation time.

\begin{table}[tb]
	\caption{Comparison of computation time between CFD and QLingNet} 
	\label{acceleratation ratio}
	\begin{center}
			\begin{tabular}{cccc}
				\hline \hline
				Airfoil        & CFD & QLingNet & acceleration ratio \\ \hline
				ah81k144       &3720.28s    &0.91s   & $4.09 \times 10^3$   \\ 
				e174           &2433.79s    &1.15s   &  $2.12 \times 10^3$\\ 
				e421           &7371.24s    &2.89s   & $2.55 \times 10^3$\\ 
				eh009          &2591.42s    &0.93s   & $2.79 \times 10^3$\\ 
				s8055          &3727.41s    &0.97s   & $3.84 \times 10^3$
				              \\ \hline \hline
			\end{tabular}
	\end{center} 
	\vspace{-1.5em}
\end{table} 

Overall, QLingNet model demonstrates good predictive accuracy when facing UIUC airfoil flow field data with varying topological structures and resolutions. 
In order to comprehensively evaluate the generalization capability of the QLingNet model, \ref{UIUC} provides additional test results concerning UIUC airfoil flow fields. 
Moreover, in \ref{Velocity and pressure}, further predictions of velocity and pressure curves for both NACA0012-CST airfoil flow field dataset and UIUC airfoil flow field dataset by the QLingNet model are presented.
\section{Conclusions}

In this work, we propose QLingNet, which has linear computational complexity and can be used for predicting flow fields of any resolution.
The CycleFC method effectively overcomes the sensitivity of traditional convolutional neural networks to flow field resolution by integrating multi-scale flow field features at the channel dimension. 
Additionally, a memory pool module is designed to store flow field resolution data from the down-sampling stage in a memory pool, ensuring accurate prediction and restoration of flow field data at different scales in the up-sampling stage.
The model's feature pyramid structure also ensures its adaptability to multi-scale flow field features. 
The test results demonstrate that QLingNet is three orders of magnitude faster than CPU-based solvers.

We provide two types of subsonic flow field datasets, NACA0012-CST and UIUC, with different geometric shapes, to test the prediction accuracy and generalization capability of the QLingNet model. 
The test results indicate that the QLingNet model achieves high fitting accuracy with the velocity and pressure fields compared to CFD calculation results, even when facing flow field data with different shapes and resolutions. 
For both velocity $u$, $v$ and pressure $p$, the prediction accuracy exceeds 99\%. 
The experiment also demonstrates that a larger amount of data significantly improves the modeling accuracy. 
In the NACA0012-CST dataset, which has more training data, the mean squared error between predicted and ground truth is on the order of $1e{-6}$, while in the UIUC dataset with fewer training data, the mean squared error remains at the order of $1e{-5}$.

Overall, the QLingNet provide strong support for engineering applications. 
In future work, we will explore how artificial intelligence techniques can be applied to simulate more complex flows rapidly, accelerate research on large-scale deep learning aerodynamics models, and expedite the industrial application of this technology.

\section *{The name of 'QLingNet'} \label{name}

The inspiration for QLingNet comes from the Qinling Mountains. 
The Qinling are a major east-west mountain range in southern Shaanxi Province, China.
The mountains mark the divide between the drainage basins of the Yangtze and Yellow River systems, providing a natural boundary between North and South China (see \url{https://en.wikipedia.org/wiki/Qinling}).

\section *{CRediT authorship contribution statement}

$\mathbf{Kuijun \quad Zuo:}$ Data curation, Formal analysis, Investigation, Methodology, Software, Validation, Visualization, Writing-original draft, Writing-review \& editing.

$\mathbf{Zhengyin \quad Ye:}$ Funding acquisition, Supervision, Resources, Project administration. 

$\mathbf{Linyang \quad Zhu:}$ Funding acquisition, Resources, Supervision.

$\mathbf{Xianxu \quad Yuan:}$ Funding acquisition, Resources, Supervision.

$\mathbf{Weiwei \quad Zhang:}$ Conceptualization, Methodology, Project administration, Supervision, Writing-review \& editing.

\section *{Acknowledgments}
This work was supported by the National Natural Science Foundation of China (Grant No. 12202470).

\section *{Data availability}

The data and code that support the findings of this study are available by contacting the corresponding author via email.

\section *{Declaration of competing interest}
The authors declare that they are have no known competing financial interests or personal relationships that could have appeared to influence the work reported in this paper.

\section *{} 

\bibliography{mybibfile}

\begin{thebibliography}{10}
\expandafter\ifx\csname url\endcsname\relax
  \def\url#1{\texttt{#1}}\fi
\expandafter\ifx\csname urlprefix\endcsname\relax\def\urlprefix{URL }\fi
\expandafter\ifx\csname href\endcsname\relax
  \def\href#1#2{#2} \def\path#1{#1}\fi

\bibitem{cutrone2024transition}
L.~Cutrone, A.~Schettino, J.~I. Cardesa, G.~Delattre, J.~G. Coder, S.~Qiang,
  M.~M. Choudhari, E.~Vogel, Transition prediction in hypersonic regime on
  complex geometries with rans-based models, in: AIAA SCITECH 2024 Forum, 2024,
  p. 0291.

\bibitem{chaiyanupong2024design}
J.~Chaiyanupong, C.~Khajorntraidet, Design and analysis of double element
  airfoil using rans, Journal of Research and Applications in Mechanical
  Engineering 12~(1) (2024).

\bibitem{SIMSEK2023114298}
O.~Simsek, H.~Islek, 2d and 3d numerical simulations of dam-break flow problem
  with rans, des, and les, Ocean Engineering 276 (2023) 114298.

\bibitem{esfahanian2024aerodynamic}
V.~Esfahanian, M.~J. Izadi, H.~Bashi, M.~Ansari, A.~Tavakoli, M.~Kordi,
  Aerodynamic shape optimization of gas turbines: a deep learning surrogate
  model approach, Structural and Multidisciplinary Optimization 67~(1) (2024)
  2.

\bibitem{salimipour2024moving}
E.~Salimipour, On the moving surface impact on flow field and aerodynamic
  performance of a thick airfoil, Ocean Engineering 291 (2024) 116504.

\bibitem{XIE2023121002}
J.~W. Hairun~Xie, M.~Zhang, Knowledge-embedded meta-learning model for lift
  coefficient prediction of airfoils, Expert Systems with Applications 233
  (2023) 121002.

\bibitem{zhong2024fast}
J.~Zhong, F.~Qu, D.~Sun, J.~Tian, T.~Wang, J.~Bai, Fast flow field prediction
  approach of supersonic inlet in wide operating range based on deep learning,
  Aerospace Science and Technology (2024) 108955.

\bibitem{shukla2024deep}
K.~Shukla, V.~Oommen, A.~Peyvan, M.~Penwarden, N.~Plewacki, L.~Bravo,
  A.~Ghoshal, R.~M. Kirby, G.~E. Karniadakis, Deep neural operators as accurate
  surrogates for shape optimization, Engineering Applications of Artificial
  Intelligence 129 (2024) 107615.

\bibitem{yetkin2024investigation}
S.~Yetkin, S.~Abuhanieh, S.~Yigit, Investigation on the abilities of different
  artificial intelligence methods to predict the aerodynamic coefficients,
  Expert Systems with Applications 237 (2024) 121324.

\bibitem{wang2024amsc}
Y.~Wang, R.~Dan, S.~Luo, L.~Sun, Q.~Wu, Y.~Li, X.~Chen, K.~Yan, X.~Ye, D.~Yu,
  Amsc-net: Anatomy and multi-label semantic consistency network for
  semi-supervised fluid segmentation in retinal oct, Expert Systems with
  Applications (2024) 123496.

\bibitem{soler2024reinforcement}
D.~Soler, O.~Mari{\~n}o, D.~Huergo, M.~de~Frutos, E.~Ferrer, Reinforcement
  learning to maximize wind turbine energy generation, Expert Systems with
  Applications 249 (2024) 123502.

\bibitem{ismael2021deep}
A.~M. Ismael, A.~{\c{S}}eng{\"u}r, Deep learning approaches for covid-19
  detection based on chest x-ray images, Expert Systems with Applications 164
  (2021) 114054.

\bibitem{zhu2024vistfc}
S.~Zhu, S.~Li, D.~Xiong, Vistfc: Vision-guided target-side future context
  learning for neural machine translation, Expert Systems with Applications
  (2024) 123411.

\bibitem{guo2016convolutional}
X.~Guo, W.~Li, F.~Iorio, Convolutional neural networks for steady flow
  approximation, in: Proceedings of the 22nd ACM SIGKDD international
  conference on knowledge discovery and data mining, 2016, pp. 481--490.

\bibitem{wu2022fast}
M.-Y. Wu, Y.~Wu, X.-Y. Yuan, Z.-H. Chen, W.-T. Wu, N.~Aubry, Fast prediction of
  flow field around airfoils based on deep convolutional neural network,
  Applied Sciences 12~(23) (2022) 12075.

\bibitem{ribeiro2020deepcfd}
M.~D. Ribeiro, A.~Rehman, S.~Ahmed, A.~Dengel, Deepcfd: Efficient steady-state
  laminar flow approximation with deep convolutional neural networks, arXiv
  preprint arXiv:2004.08826 (2020).

\bibitem{hu2022mesh}
J.-W. Hu, W.-W. Zhang, Mesh-conv: Convolution operator with mesh resolution
  independence for flow field modeling, Journal of Computational Physics 452
  (2022) 110896.

\bibitem{duru2021cnnfoil}
C.~Duru, H.~Alemdar, {\"O}.~U. Baran, Cnnfoil: Convolutional encoder decoder
  modeling for pressure fields around airfoils, Neural Computing and
  Applications 33~(12) (2021) 6835--6849.

\bibitem{duru2022deep}
C.~Duru, H.~Alemdar, O.~U. Baran, A deep learning approach for the transonic
  flow field predictions around airfoils, Computers \& Fluids 236 (2022)
  105312.

\bibitem{sun2021deep}
D.~Sun, Z.~Wang, F.~Qu, J.~Bai, A deep learning based prediction approach for
  the supercritical airfoil at transonic speeds, Physics of Fluids 33~(8)
  (2021).

\bibitem{leer2021fast}
M.~Leer, A.~Kempf, Fast flow field estimation for various applications with a
  universally applicable machine learning concept, Flow, Turbulence and
  Combustion 107 (2021) 175--200.

\bibitem{haizhou2022generative}
W.~Haizhou, L.~Xuejun, A.~Wei, L.~Hongqiang, A generative deep learning
  framework for airfoil flow field prediction with sparse data, Chinese Journal
  of Aeronautics 35~(1) (2022) 470--484.

\bibitem{wu2020deep}
H.~Wu, X.~Liu, W.~An, S.~Chen, H.~Lyu, A deep learning approach for efficiently
  and accurately evaluating the flow field of supercritical airfoils, Computers
  \& Fluids 198 (2020) 104393.

\bibitem{wang2023general}
Z.~Wang, X.~Liu, J.~Yu, H.~Wu, H.~Lyu, A general deep transfer learning
  framework for predicting the flow field of airfoils with small data,
  Computers \& Fluids 251 (2023) 105738.

\bibitem{yang2022amgnet}
Z.~Yang, Y.~Dong, X.~Deng, L.~Zhang, Amgnet: Multi-scale graph neural networks
  for flow field prediction, Connection Science 34~(1) (2022) 2500--2519.

\bibitem{ogoke2021graph}
F.~Ogoke, K.~Meidani, A.~Hashemi, A.~B. Farimani, Graph convolutional networks
  applied to unstructured flow field data, Machine Learning: Science and
  Technology 2~(4) (2021) 045020.

\bibitem{li2022integrated}
J.~Li, T.~Liu, Y.~Wang, Y.~Xie, Integrated graph deep learning framework for
  flow field reconstruction and performance prediction of turbomachinery,
  Energy 254 (2022) 124440.

\bibitem{chen2024pointgpt}
G.~Chen, M.~Wang, Y.~Yang, K.~Yu, L.~Yuan, Y.~Yue, Pointgpt: Auto-regressively
  generative pre-training from point clouds, Advances in Neural Information
  Processing Systems 36 (2024).

\bibitem{abbas2022geometrical}
A.~Abbas, A.~Rafiee, M.~Haase, A.~Malcolm, Geometrical deep learning for
  performance prediction of high-speed craft, Ocean Engineering 258 (2022)
  111716.

\bibitem{jiang2023transcfd}
J.~Jiang, G.~Li, Y.~Jiang, L.~Zhang, X.~Deng, Transcfd: A transformer-based
  decoder for flow field prediction, Engineering Applications of Artificial
  Intelligence 123 (2023) 106340.

\bibitem{hemmasian2023reduced}
A.~Hemmasian, A.~Barati~Farimani, Reduced-order modeling of fluid flows with
  transformers, Physics of Fluids 35~(5) (2023).

\bibitem{deng2023prediction}
Z.~Deng, J.~Wang, H.~Liu, H.~Xie, B.~Li, M.~Zhang, T.~Jia, Y.~Zhang, Z.~Wang,
  B.~Dong, Prediction of transonic flow over supercritical airfoils using
  geometric-encoding and deep-learning strategies, arXiv preprint
  arXiv:2303.03695 (2023).

\bibitem{zuo2023fast}
K.~Zuo, Z.~Ye, W.~Zhang, X.~Yuan, L.~Zhu, Fast aerodynamics prediction of
  laminar airfoils based on deep attention network, Physics of Fluids 35~(3)
  (2023).

\bibitem{chen2107cyclemlp}
S.~Chen, E.~Xie, C.~Ge, R.~Chen, D.~Liang, P.~Luo, Cyclemlp: A mlp-like
  architecture for dense prediction. arxiv 2021, arXiv preprint
  arXiv:2107.10224.

\bibitem{swannet2024towards}
K.~Swannet, C.~Varriale, A.~K. Doan, Towards universal parameterization: Using
  variational autoencoders to parameterize airfoils, in: AIAA SCITECH 2024
  Forum, 2024, p. 0686.

\bibitem{lane2010inverse}
K.~Lane, D.~Marshall, Inverse airfoil design utilizing cst parameterization,
  in: 48th AIAA Aerospace Sciences Meeting Including the New Horizons Forum and
  Aerospace Exposition, 2010, p. 1228.

\bibitem{zhao2020design}
Z.~Zhao, L.~He, X.-y. HE, Design of general cfd software phenglei, Computer
  Engineering \& Science 42~(02) (2020) 210.

\bibitem{zuo2024fast}
K.~Zuo, Z.~Ye, S.~Bu, X.~Yuan, W.~Zhang, Fast simulation of airfoil flow field
  via deep neural network, arXiv preprint arXiv:2312.04289 (2023).

\end{thebibliography}

\newpage
\setcounter{figure}{0} 
\appendix 

\section{NACA0012-CST dataset test results} \label{NACA0012 CST}

\begin{figure*}[!h]
	\begin{center}
		\includegraphics[width=1 \linewidth]{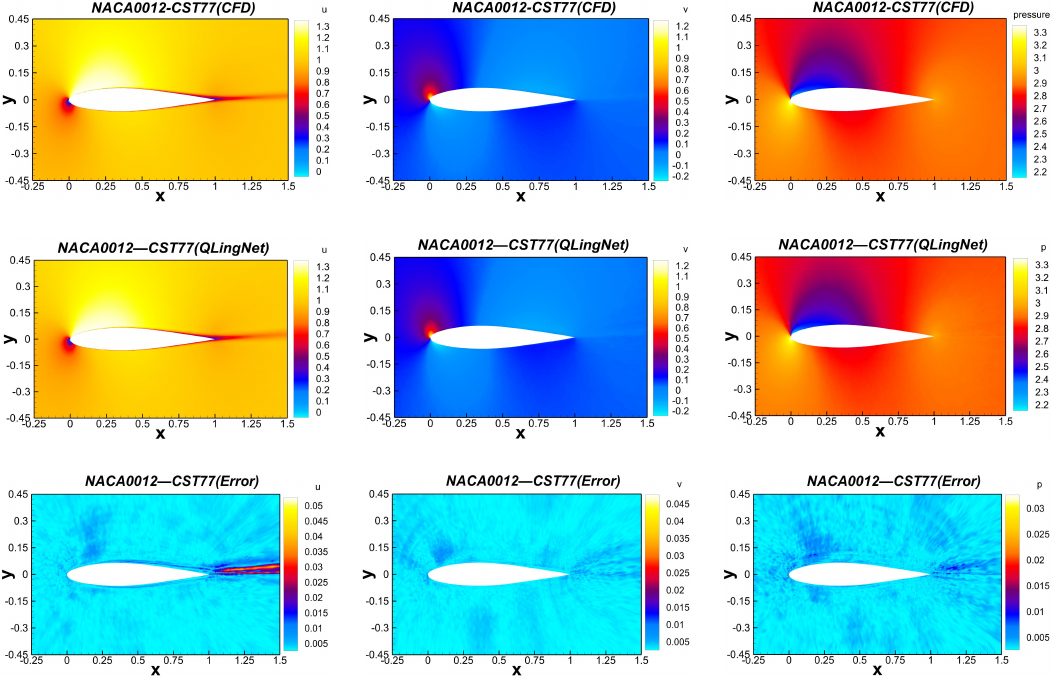}
	\end{center}  \vspace{-2mm}  
	\caption{{{NACA0012-CST77 airfoil flow field CFD calculation results, QLingNet prediction results, and corresponding absolute error plot.}
	}} \label{naca0012_cst_77} 
\end{figure*}

\begin{figure*}[!h]
	\begin{center}
		\includegraphics[width=1 \linewidth]{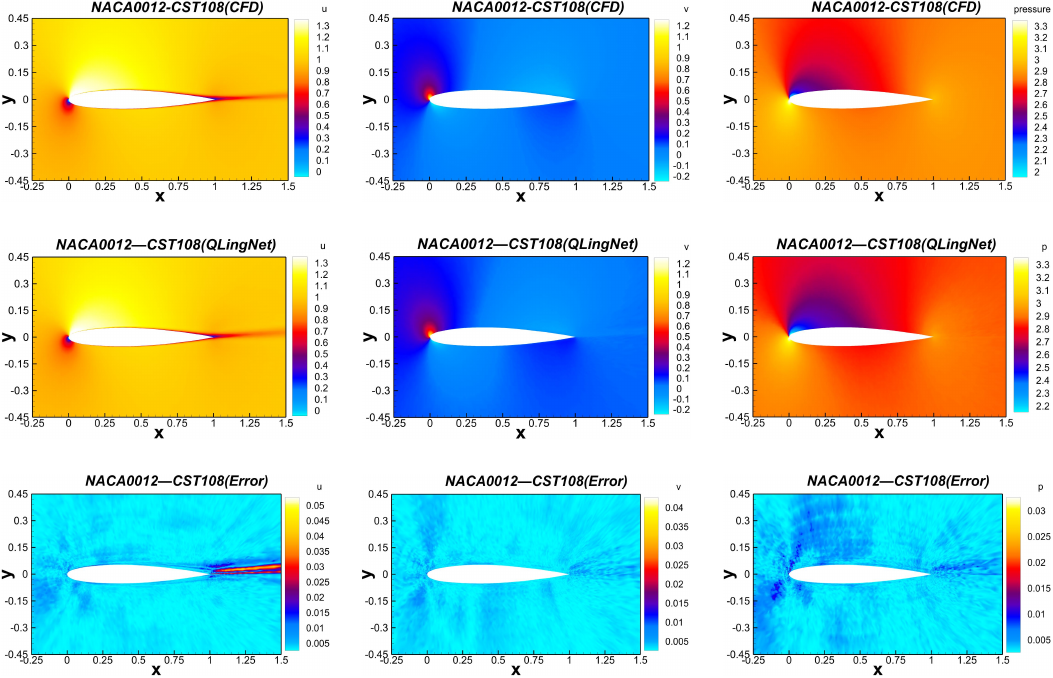}
	\end{center}  \vspace{-2mm}  
	\caption{{{NACA0012-CST108 airfoil flow field CFD calculation results, QLingNet prediction results, and corresponding absolute error plot.}
	}} \label{naca0012_cst_108} 
\end{figure*}

\begin{figure*}[!h]
	\begin{center}
		\includegraphics[width=1 \linewidth]{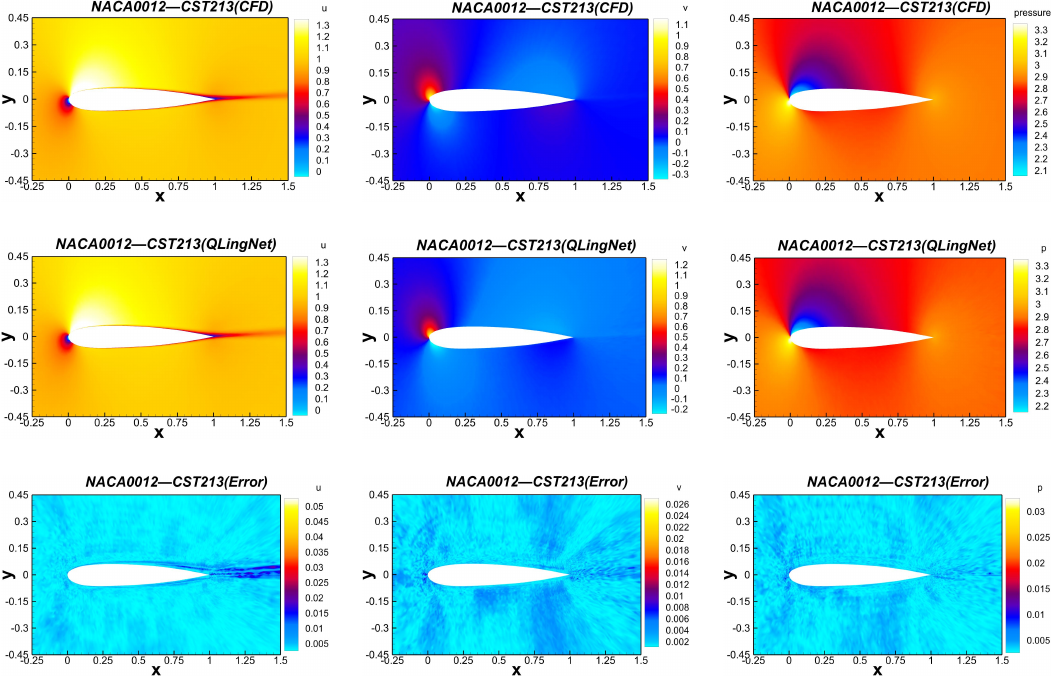}
	\end{center}  \vspace{-2mm}  
	\caption{{{NACA0012-CST213 airfoil flow field CFD calculation results, QLingNet prediction results, and corresponding absolute error plot.}
	}} \label{naca0012_cst_213} 
\end{figure*}

\begin{figure*}[!h]
	\begin{center}
		\includegraphics[width=1 \linewidth]{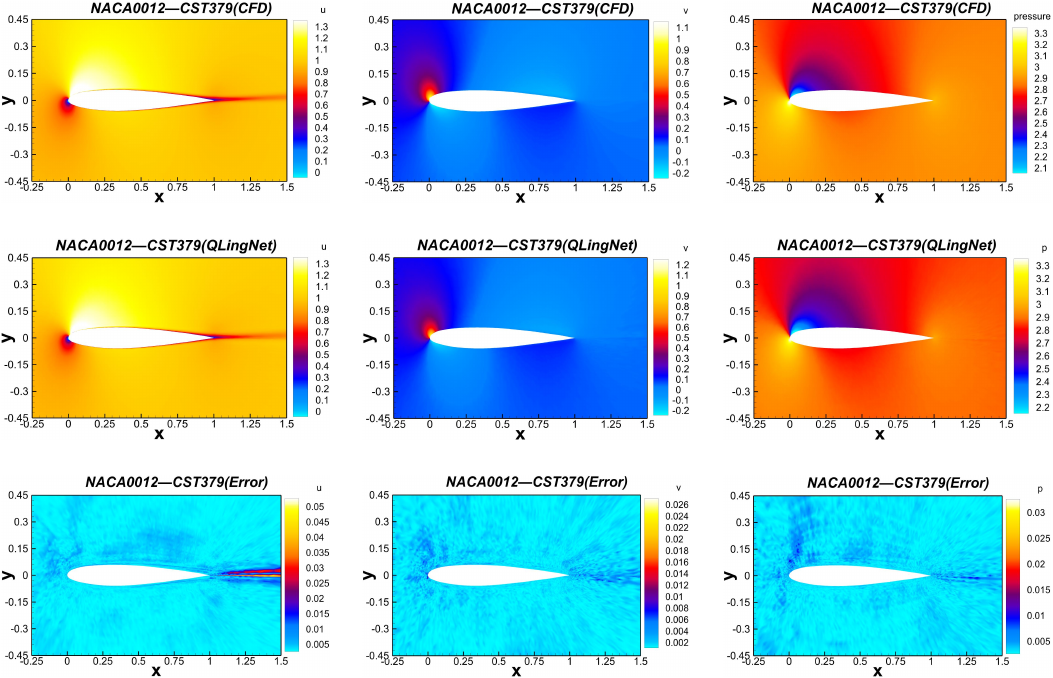}
	\end{center}  \vspace{-2mm}  
	\caption{{{NACA0012-CST379 airfoil flow field CFD calculation results, QLingNet prediction results, and corresponding absolute error plot.}
	}} \label{naca0012_cst_379} 
\end{figure*}

\clearpage
\setcounter{figure}{0} 
\section{UIUC dataset test results} \label{UIUC}

\begin{figure*}[!h]
	\begin{center}
		\includegraphics[width=0.9 \linewidth]{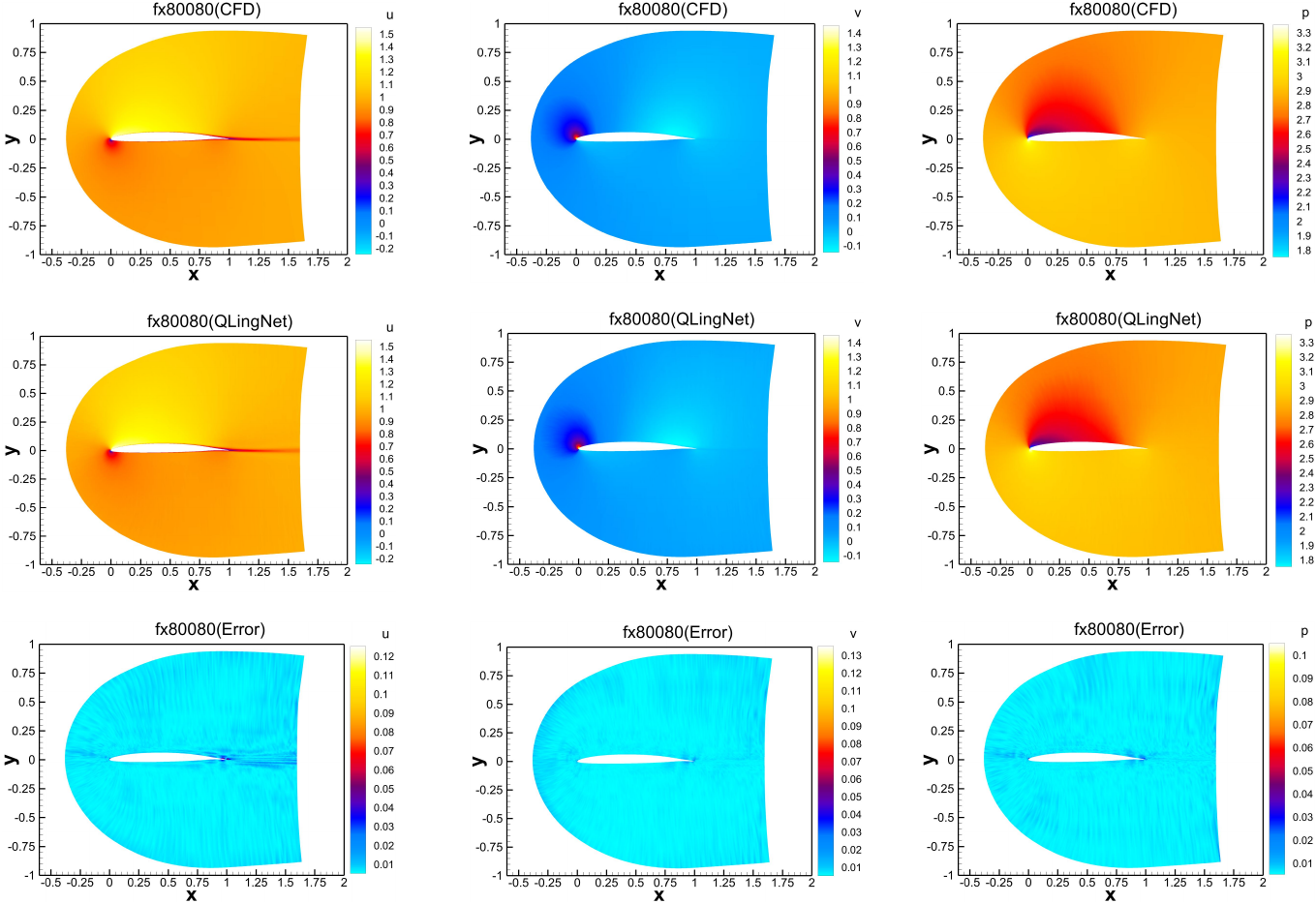}
	\end{center}  \vspace{-2mm}  
	\caption{{{fx80080 airfoil flow field CFD calculation results, QLingNet prediction results, and corresponding absolute error plot.}
	}} \label{fx80080} 
\end{figure*}

\begin{figure*}[!h]
	\begin{center}
		\includegraphics[width=0.9\linewidth]{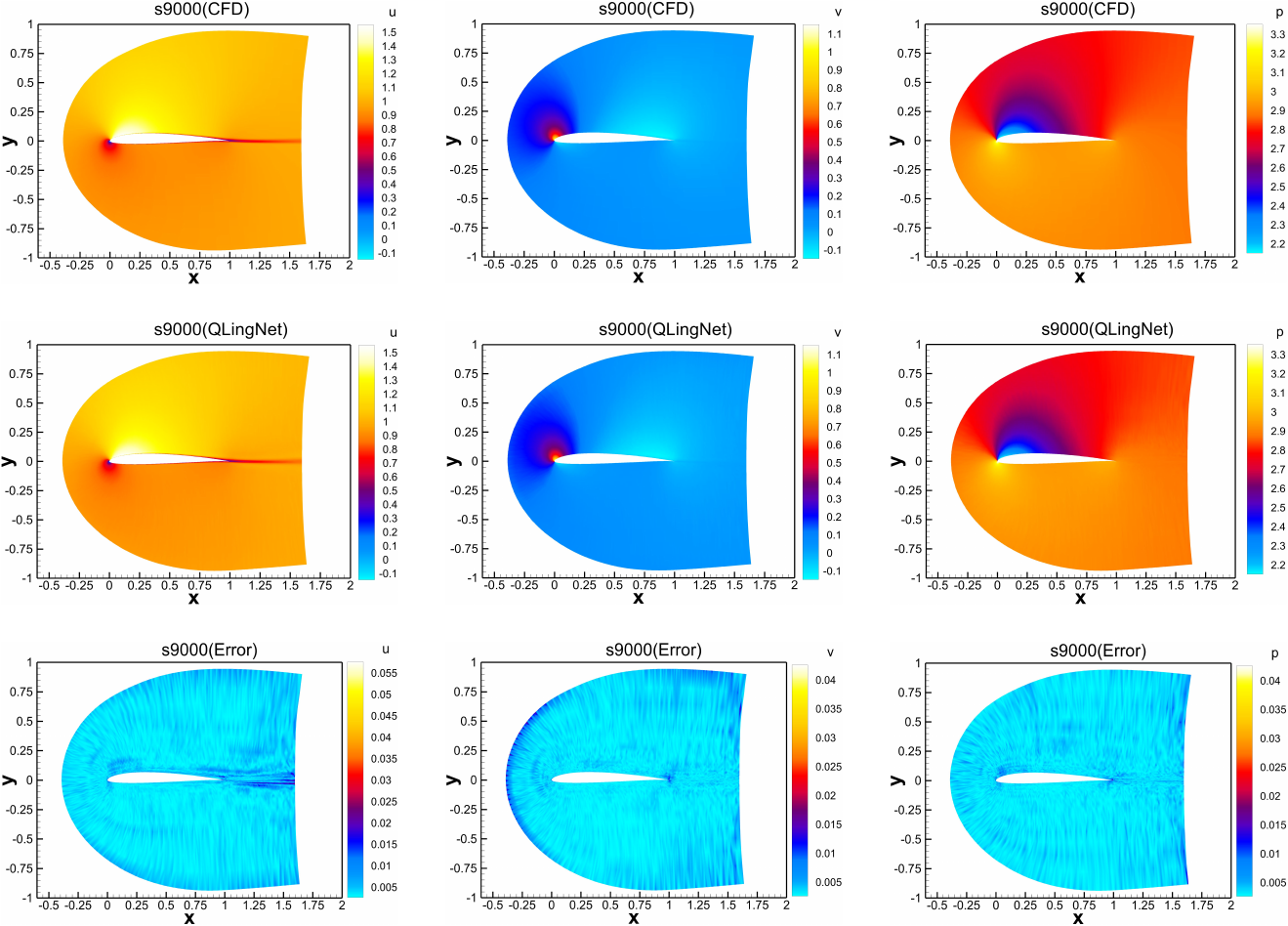}
	\end{center}  \vspace{-2mm}  
	\caption{{{s9000 airfoil flow field CFD calculation results, QLingNet prediction results, and corresponding absolute error plot.}
	}} \label{s9000} 
\end{figure*}

\setcounter{figure}{0} 
\setcounter{table}{0} 
\section{Fitting curves for velocity and pressure} \label{Velocity and pressure}

\begin{figure*}[!h]
	\begin{center}
		\includegraphics[width=1 \linewidth]{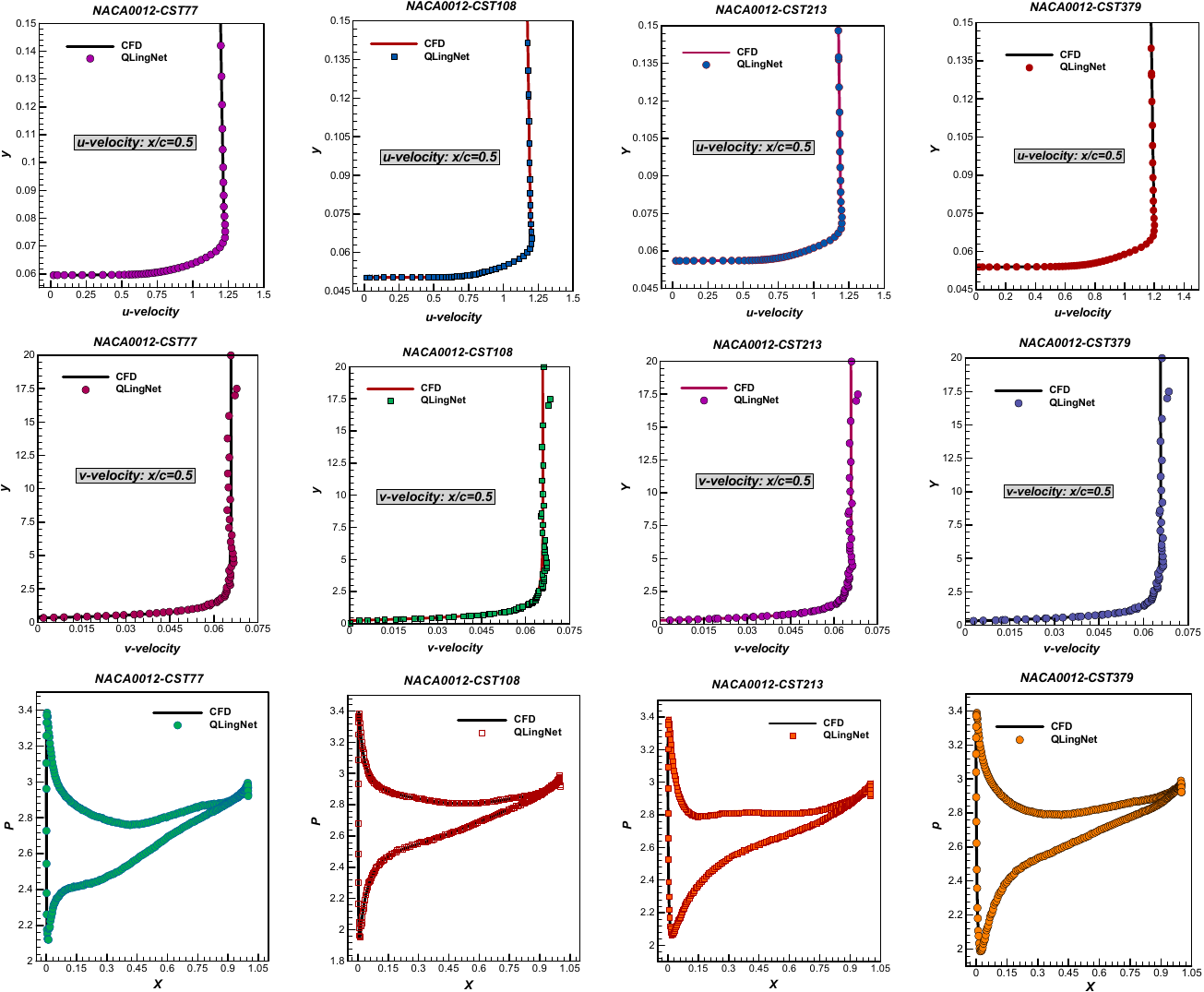}
	\end{center}  \vspace{-2mm}  
	\caption{{{Velocity and pressure fitting curves for four different test cases in the NACA0012-CST dataset.}   
	}} \label{naca0012_cst_velocity} 
\end{figure*}

\begin{figure*}[!h]
	\begin{center}
		\includegraphics[width=1 \linewidth]{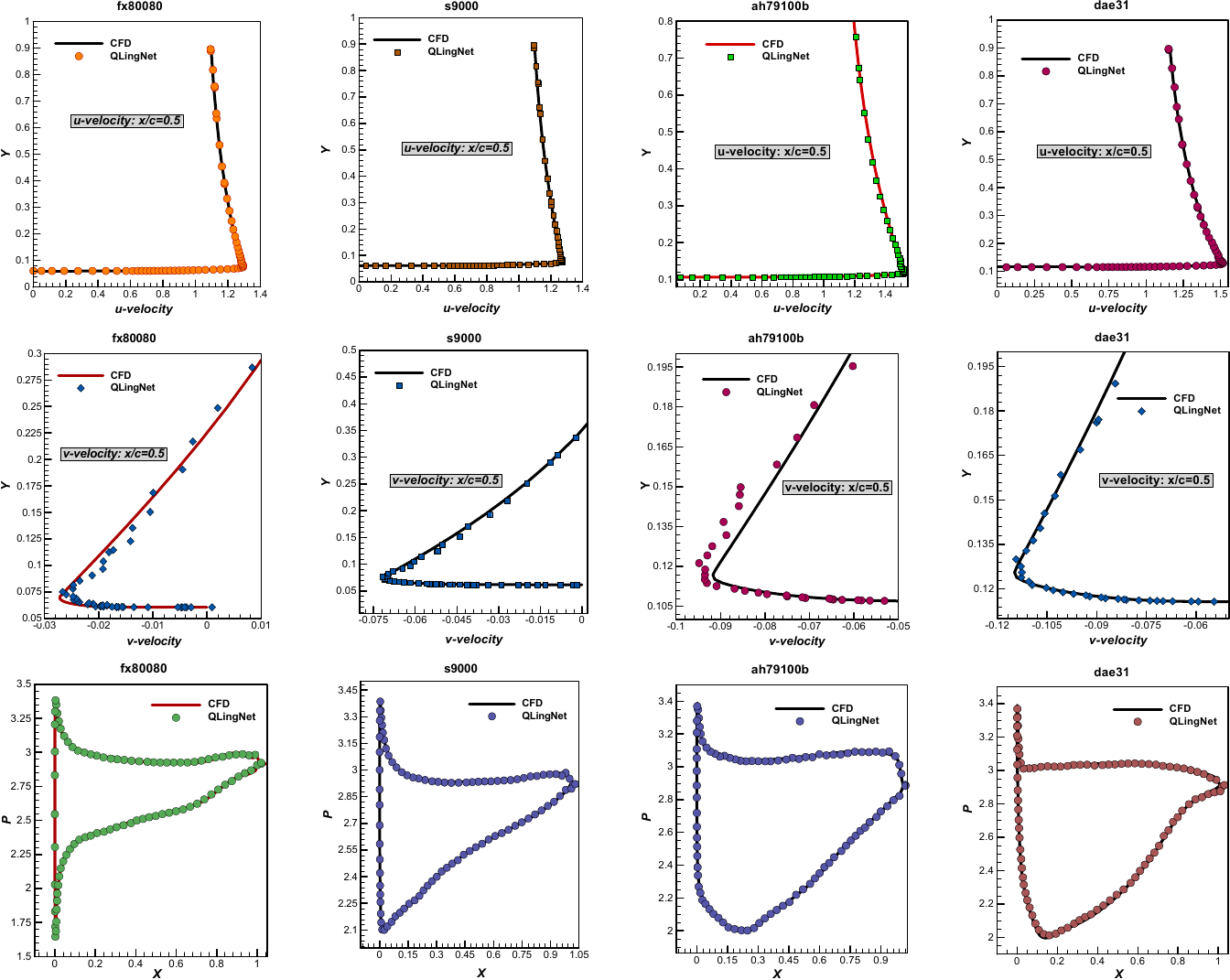}
	\end{center}  \vspace{-2mm}  
	\caption{{{Velocity and pressure fitting curves for four different test cases in the UIUC dataset.}   
	}} \label{uiuc_velocity} 
\end{figure*}

\clearpage
Use Mean Squared Error (MSE), Mean Absolute Error (MAE), and Mean Absolute Percentage Error (MAPE) to evaluate the prediction accuracy of the velocity and pressure curves in Figure \ref{naca0012_cst_velocity} and Figure \ref{uiuc_velocity}, as well as the prediction accuracy of the global pressure field.
MSE represents the average squared difference between the predicted values and the true values for all samples. 
Its calculation formula is as follows:

\begin{equation}
	\mathrm{MSE}=\frac1n\sum_{i=1}^n(x_i-\hat{x_i})^2
\end{equation}
Where $n$ denotes the number of samples, $x_i$ represents the ground truth of the $i$-th sample, and $\hat{x}_i$ represents the predicted value of the $i$-th sample.
MAE represents the average absolute difference between the predicted values and the true values for all samples. 
Its calculation formula is as follows:

\begin{equation}
	\mathrm{MAE}=\frac{1}{n}\sum_{i=1}^{n}|y_{i}-\hat{y}_{i}|
\end{equation}

The MAPE represents the average absolute value of the ratio of the prediction error to the true value for all samples, expressed as a percentage. 
Its calculation formula is as follows:

\begin{equation}
	\mathrm{MAPE}=\frac1n\sum_{i=1}^n\left|\frac{z_i-\hat{z}_i}{z_i}\right|\times100\%
\end{equation}

In Table \ref{naca0012cst_uiuc_loss table}, MSE, MAE, and MAPE are denoted by symbols $\delta$, $\epsilon$ and $\varrho$, respectively. 
$\digamma$ represents the error of the whole-domain pressure field, and 
$\varsigma$ represents the error of the curve in Figure \ref{naca0012_cst_velocity} and \ref{uiuc_velocity}.
For example, $U_{\delta, \varsigma}$ represents the Mean Squared Error of the velocity curve $u$.
$P_{\varrho, \digamma}$ represents the MAPE of the whole-domain pressure field.

\begin{table}[tb]
\caption{Error analysis for various test cases} 
\label{naca0012cst_uiuc_loss table}
\begin{center}
		\begin{tabular}{ccccccc}
			\hline
			Airfoil         & $U_{\varrho, \varsigma}$ & $V_{\varrho, \varsigma}$ & $P_{\varrho, \varsigma}$ & $P_{\varrho, \digamma}$ & $P_{\delta, \digamma}$ & $P_{\epsilon, \digamma}$ \\ \hline
			NACA0012-CST77  & $0.358\%$ & $2.775\%$ & $0.120\%$ & $0.057\%$    & $5.496 \times 10^{-6}$      & $1.606 \times 10^{-3}$ \\
			NACA0012-CST108 & $0.443\%$ & $8.640\%$ & $0.158\%$ &   $0.066\%$  & $7.443 \times 10^{-6}$      & $1.842 \times 10^{-3}$ \\
			NACA0012-CST213 & $0.326\%$ & $3.369\%$ & $0.118\%$ & $0.059\%$    & $5.415 \times 10^{-6}$      & $1.655 \times 10^{-3}$ \\
			NACA0012-CST379 & $0.275\%$ & $2.588\%$ & $0.095\%$ & $0.048\%$    & $3.948 \times 10^{-6}$     & $1.345 \times 10^{-3}$ \\
			fx80080         & $0.508\%$ & $13.243\%$ & $0.281\%$ & $0.156\%$    & $6.639 \times 10^{-5}$      & $4.011 \times 10^{-3}$ \\
			s9000           & $0.450\%$ & $4.611\%$ & $0.180\%$ & $0.108\%$    & $1.741 \times 10^{-5}$      & $2.921 \times 10^{-3}$ \\
			ah79100b        & $0.778\%$ & $10.648\%$ & $0.316\%$ & $0.184\%$    & $4.646 \times 10^{-5}$      &$5 \times 10^{-3}$ \\
			dae31           & $0.657\%$ & $9.293\%$ & $0.376\%$ & $0.222\%$    & $7.517 \times 10^{-5}$      & $5.808 \times 10^{-3}$ \\ \hline
		\end{tabular}
\end{center} 
\vspace{-1.5em}
\end{table} 

\end{CJK}
\end{document}